# Pulsar timing array harmonic analysis and source angular correlations

Bruce Allen[*]

*Max Planck Institute for Gravitational Physics* (Albert Einstein Institute), *Leibniz Universität Hannover, Callinstrasse 38, D-30167, Hannover, Germany*

Gravitational waves (GWs) influence the arrival times of radio signals coming from pulsars. Here, we investigate the harmonic space approach to describing a pulsar's response to GWs. We derive and discuss the "diagonalized form" of the response, which is a sum of spin-2-weighted spherical harmonics of the GW direction multiplied by normal (spin-weight 0) spherical harmonics of the pulsar direction. We show how this allows many useful objects, for example, the Hellings and Downs two-point function, to be easily calculated. The approach also provides a clear description of the gauge dependence. We then employ this harmonic approach to model the effects of angular correlations in the sky locations of GW sources (sometimes called "statistical isotropy"). To do this, we construct ensembles made up of many Gaussian subensembles. While each of the individual subensembles breaks rotational invariance, the full ensemble is rotationally invariant. Using harmonic techniques, we compute the cosmic covariance and the total covariance of the Hellings and Downs correlation in these models. The results may be used to assess the impact of angular source correlations on the Hellings and Downs correlation, and for optimal reconstruction of the Hellings and Downs curve in models where GW sources have correlated sky locations.

## I. INTRODUCTION

There is a considerable literature on the topic of pulsar timing arrays (PTAs), which may be on the verge of making five-sigma detections of nHz gravitational waves (GWs) [1–4]. PTAs rely on the effect that GWs have on shifting the arrival times of radio pulses. An introductory discussion of how they work can be found in [5].

As shown in [5] [Sec. 2.2, Eq. (14)], the shift in arrival times is due to the Sachs-Wolfe effect [6], which also creates temperature fluctuations of the (electromagnetic) cosmic background radiation (CBR). So it is not surprising that there is a considerable literature which applies tools and techniques drawn from CBR to PTAs [7–10]. Here, we use the term "harmonic analysis" for this approach, which describes the response of PTA pulsars in terms of spherical harmonic functions on the two-dimensional sphere.

This paper presents the most important of these tools and results from a physical perspective, and illustrates how they may be used to describe the response of PTAs. While much of this can be found in the corresponding specialist literature [10–25], and the seminal paper on the topic [8] is a decade old, we hope to offer some fresh insights as well as a few new results. A brief outline of the paper, including links to key equations, follows.

Our analysis assumes that pulsars are perfect clocks: the response of a pulsar to a GW can be described as a "redshift" (or "blueshift") of the clock frequency. In Sec. II, we review the way in which pulsar redshift $Z$ responds to the GW amplitude at Earth. The response $F(\Omega, \Omega_p)$ is a function (2.2) of the direction $\Omega$ of GW propagation and of the direction $\Omega_p$ to the pulsar. Here, $\Omega$ and $\Omega_p$ are (coordinates of) points on the unit two-sphere, and $\hat{\Omega}$ and $\hat{\Omega}_p$ are the corresponding unit vectors.

The function $F$ is complex, with real and imaginary parts that describe, respectively, the response to plus- and cross-polarized GWs. While the magnitude (2.7) of this response depends only upon the angle between the direction of GW propagation and the direction to the pulsar in the sky, the phase of the response has a more complicated dependence on these two positions. We write $F$ in diagonal form (2.10) as a sum of spin-weight 0 harmonic functions of $\Omega_p$ and spin-weight 2 harmonic functions of $\Omega$. This reflects the spin-two nature of GWs, correctly incorporating both the magnitude and the phase of the response.

The remainder of the paper exploits this diagonal form. In Sec. III, we use it to derive the harmonic-space form of the Hellings and Downs (HD) curve (3.3), by averaging over source directions $\Omega$, which was the original approach employed in 1983 by Hellings and Downs [26]. In Sec. IV, we derive a simple formula that can be used to "pulsar average" any function $Q(\Omega_p, \Omega_q)$ of the sky directions to pulsars $p$ and $q$, as illustrated in Fig. 5 of [27]. This produces a function $Q(\gamma)$ of a single variable $\gamma$, which is the angle between the directions to two pulsars (4.4). In Sec. V, we again use the harmonic decomposition of $F$ to derive the HD curve (5.1), but this time as the pulsar average of the correlation for a single GW source. This equality between the source direction average for a single pulsar pair and the pulsar average for a single source direction was first demonstrated, highlighted, and discussed in [28]. In comparison with the original approach employed by Hellings and Downs [26], we believe that this is a "better way" to define and to think about the HD curve.

---

[*]Contact author: bruce.allen@aei.mpg.de

In Sec. VI, we turn attention to the HD two-point function $\mu(\gamma, \Omega, \Omega')$, which is defined by (6.1) and was first computed in [27] (Appendix G). This is the average correlation of a pair of pulsars separated by angle $\gamma$, for GWs from sources radiating in directions $\Omega$ and $\Omega'$. We compute it using the pulsar averaging recipe from Sec. IV, showing that the magnitude of this quantity depends only upon $\gamma$ and upon the angle $\beta$ between the two sources (6.8). We obtain a beautiful new harmonic form (6.9) for the two-point function, as a sum of products of Legendre and Jacobi polynomials, which is used later in the paper to study the cosmic covariance and variance.

Starting in Sec. VII, we employ Gaussian statistical ensembles of GW sources. As shown by the central limit theorem, these describe GWs which are produced by the incoherent sum of many weak sources. Working in a circular polarization basis, we first compute the cosmic covariance for the standard ensemble of unpolarized sources [fully defined by first and second moments (7.4)] demonstrating explicitly how the phase of $\mu(\gamma, \Omega, \Omega')$ cancels out. Then, in Sec. VIII, we exploit the harmonic form of the two-point function found earlier, to obtain explicit harmonic decompositions of the cosmic covariance (8.4) and variance (8.5).

In Sec. IX, we consider cosmological ensembles where the source sky locations have nonzero angular covariance (they are not a Poisson process [29,30] in the limit of an infinite density of vanishingly weak sources.) This ensemble is sometimes called "statistically isotropic" in contrast with the standard Gaussian ensemble, which is called "purely isotropic." (These names are misleading: while both ensembles are isotropic, a typical representative universe drawn from *either* ensemble breaks rotational invariance, and thus is not isotropic.)

In this paper, a statistical ensemble is a (possibly infinite) collection of representative universes containing GWs. Each of these representative universes is called a "realization": it contains a specific set of GW sources, with specific properties (frequencies, amplitudes, …) at specific sky locations. An ensemble is rotationally invariant if and only if, for every realization, it contains all rotated (about the Earth/origin) versions of that realization.

Our approach is to build statistical ensembles composed of Gaussian subensembles (but note that the resulting full ensemble is *not* Gaussian [31]). Each individual Gaussian subensemble has preferred directions, defined by a function $\psi(\Omega)$ associated with that subensemble, which breaks rotational invariance. Nevertheless, the full ensemble maintains rotational invariance, because for any Gaussian subensemble that it contains with a given $\psi$, it contains all other Gaussian subensembles described by rotated versions of $\psi$. This construction is detailed in Sec. IX A. It has been used before, for example in [32–34], but without this explicit description [35].

This approach enables the study of GW source models where the (sky) locations of the sources have nontrivial angular covariance. Such correlations arise for any ensemble constructed from a finite number of GW sources at discrete sky locations [36]. This explicit construction provides a sound basis for similar "statistical isotropy" calculations which appear in the literature [32–34] but whose justification is problematic [37]. The correlation function $C(\Omega, \Omega') + 1 = \langle \psi(\Omega)\psi(\Omega')\rangle_\psi$ that describes correlations among GW source locations is an average over all subensembles (9.3), and only depends upon the dot product $\hat{\Omega} \cdot \hat{\Omega}'$ of the directions to the two sources. The coefficients $C_L$ of the Legendre-polynomial decomposition (9.4) of $C(\hat{\Omega} \cdot \hat{\Omega}')$ characterize the type/degree of the angular correlations.

In Sec. IX B, we compute the cosmic variance and covariance for this correlated-in-angle ensemble. (It would be logical to begin with the total variance/covariance, but that is more complicated, so we do it after.) The averages within a given Gaussian subensemble lead to intermediate results such as (9.6) and (9.8) that are exactly as for the standard case, except that they contain factors of $\psi(\Omega)$. Averaging over the full ensemble then introduces the function $C$. By employing the harmonic decomposition of the two-point function, a simple result for the cosmic covariance (9.13) is obtained.

In Sec. IX C, we return to the calculation of the total covariance. Again, we compute the mean (9.21) and correlation (9.22) of the HD correlation within a given Gaussian subensemble. Again, these are identical to standard results, apart from containing factors of $\psi(\Omega)$. In Sec. IX D, we then carry out the averages over subensembles, to obtain the covariance $\mathcal{C}_{pq,rs}$ (9.29) of the HD correlation for the full ensemble. The latter is determined by a function $D_{pq,rs}$ (9.27) of four pulsar directions $\Omega_p$, $\Omega_q$, $\Omega_r$, and $\Omega_s$. (Others have also investigated this quantity, see Appendix E of [8] and citations therein.) We derive a rotationally invariant form for $D$, given by (9.37) and (9.38). This could be used for optimal reconstructions of the HD correlation curve [38] that take account of correlations among source sky positions.

In Sec. IX E, we compute the total variance for the ensemble of correlated sources, which is the diagonal part $C_{pq,pq}$ of the covariance. From symmetry, this only depends upon the angle $\gamma$ between the directions to $p$ and $q$, so it is unaffected by pulsar averaging. By carrying out a pulsar average, the total variance can be expressed as a sum of Legendre polynomials (9.43), where the coefficients for a given $C_L$ are explicitly given in (9.47). This is followed by a brief conclusion.

The appendixes contain technical details. In Appendix A we provide key formulas for spin-weighted spherical harmonics. In Appendix B we derive the diagonal form of the HD response, and in Appendix C we compute two-point functions for the four different combinations of linear polarizations (C6).

## II. RESPONSE OF A PULSAR TO A GRAVITATIONAL WAVE

It is common to talk about physical effects in terms of fields and particles, for example, the influence of an electric field on an electron. PTAs can be described in similar terms: the influence of GWs on the arrival time of pulses [5,39]. The pulsars may be thought of as ideal clocks, and the influence of GWs is to reduce or increase their tick rates. This can be quantified as a time-dependent redshift or blueshift, which is the time derivative of the timing residual. These same quantities, redshift and blueshift, are also used to describe temperature fluctuations in the CBR.

Note that for both PTAs and CBR, the use of "red" and "blue" is a historical misnomer. The clock frequency for PTAs is hundreds of Hz, and the CBR consists of infrared radiation. Both frequencies are well below the visible part of the spectrum.

We begin with the response of a pulsar to a GW traveling in direction $\Omega$, where this symbol indicates a pair of angles $\theta, \phi$ in usual spherical polar coordinates. The corresponding unit-length vector from the origin (Earth) is denoted

$$\hat{\Omega} \equiv \cos\phi \sin\theta \, \hat{x} + \sin\phi \sin\theta \, \hat{y} + \cos\theta \, \hat{z}. \quad (2.1)$$

Note that, as seen from Earth, the sky direction to the very distant GW source is $-\hat{\Omega}$.

The sky coordinates of the pulsar are $\Omega_p$. A unit vector from Earth to the pulsar has the same components as in (2.1) but with coordinates $\theta_p, \phi_p$. Note that we typically label or index pulsars with the subscripts $p$, $q$, $r$, and $s$.

The (redshift) response of the pulsar consists of an "Earth term" and a "pulsar term." Until the distance to pulsars is known to about light-year precision, the latter cannot be measured, so in this paper we mostly ignore it [40]. The Earth term (for a unit-amplitude, circularly polarized GW) is

$$F(\Omega, \Omega_p) \equiv \frac{1}{2} \frac{(\hat{\Omega}_p \cdot (\hat{m}(\Omega) + i\hat{n}(\Omega)))^2}{1 + \hat{\Omega} \cdot \hat{\Omega}_p}, \quad (2.2)$$

where $\hat{m}$ and $\hat{n}$ are a pair of unit length vectors which are (a) perpendicular to the GW direction $\hat{\Omega}$ and (b) perpendicular to each other:

$$\hat{m}(\Omega) \equiv \cos\phi \cos\theta \, \hat{x} + \sin\phi \cos\theta \, \hat{y} - \sin\theta \, \hat{z},$$
$$\hat{n}(\Omega) \equiv -\sin\phi \, \hat{x} + \cos\phi \, \hat{y}. \quad (2.3)$$

This choice of $\hat{m}$ and $\hat{n}$ is important for what follows, but note that it is inconsistent with much of the literature.

To describe the GW, we could have made different choices for $\hat{m}$ and $\hat{n}$, rotating them in the $\hat{m}$-$\hat{n}$ plane through an angle which is an *arbitrary* function of $\Omega = \theta, \phi$. The choice we have made "fixes the gauge," ensuring that $F(\Omega, \Omega_p)$ is only a function of $\theta$, $\theta_p$, and $\phi - \phi_p$. For other choices of gauge, $F(\Omega, \Omega_p)$ might also have depended upon $\phi + \phi_p$.

The two GW polarizations are usually labeled "+" and "×," corresponding to polarization tensors

$$e^+_{ab}(\Omega) \equiv \hat{m}_a(\Omega)\hat{m}_b(\Omega) - \hat{n}_a(\Omega)\hat{n}_b(\Omega), \quad (2.4)$$

$$e^\times_{ab}(\Omega) \equiv \hat{m}_a(\Omega)\hat{n}_b(\Omega) + \hat{n}_a(\Omega)\hat{m}_b(\Omega). \quad (2.5)$$

Here, the coordinate indices $a$ and $b$ denote $x$, $y$, and $z$ components.

In what follows, we adopt the Einstein summation convention: if a coordinate index appears twice in a given term, then for that term the repeated index should be summed over the three coordinates. Some examples: $\hat{m}_a\hat{m}_a \equiv \sum_{a \in x,y,z} \hat{m}_a\hat{m}_a = \hat{m}\cdot\hat{m} = 1$; $\hat{m}_a\hat{n}_a = 0$; $e^+_{ab}e^+_{ab} = e^\times_{ab}e^\times_{ab} = 2$; $e^+_{ab}e^\times_{ab} = 0$.

One can see from inspection of (2.2) that the real part of the numerator of $F$ is $\hat{\Omega}^a_p \hat{\Omega}^b_p e^+_{ab}(\Omega)$ and that the imaginary part is $\hat{\Omega}^a_p \hat{\Omega}^b_p e^\times_{ab}(\Omega)$. Hence, the real part of $F$ is the redshift produced by a plus-polarized GW with unit amplitude at Earth at that moment in time, and the imaginary part of $F$ is the redshift produced by a cross-polarized GW: $F(\Omega, \Omega_p) = F_+(\Omega, \Omega_p) + iF_\times(\Omega, \Omega_p)$. Thus, $F$ is the instantaneous redshift response of a pulsar to a circularly polarized GW of unit strain amplitude at Earth, where "circularly polarized" means a polarization tensor $e^+_{ab} + ie^\times_{ab}$.

It is easy to see that the modulus of the response $|F(\Omega, \Omega_p)|$ only depends upon the angle between $\hat{\Omega}$ and $\hat{\Omega}_p$. The modulus of the numerator of (2.2) is

$$\begin{aligned}\left|(\hat{\Omega}_p \cdot (\hat{m} + i\hat{n}))^2\right| &= |\hat{\Omega}_p \cdot (\hat{m} + i\hat{n})|^2\\ &= \hat{\Omega}^a_p \hat{\Omega}^b_p (\hat{m}_a\hat{m}_b + \hat{n}_a\hat{n}_b)\\ &= \hat{\Omega}^a_p \hat{\Omega}^b_p (\delta_{ab} - \hat{\Omega}_a\hat{\Omega}_b)\\ &= 1 - (\hat{\Omega} \cdot \hat{\Omega}_p)^2. \end{aligned} \quad (2.6)$$

The third equality follows since $\hat{\Omega}$, $\hat{m}$, and $\hat{n}$ form an orthonormal basis, so $\delta_{ab} = \hat{m}_a\hat{m}_b + \hat{n}_a\hat{n}_b + \hat{\Omega}_a\hat{\Omega}_b$ is the $3 \times 3$ identity matrix or the Kronecker delta. The fourth equality is because $\hat{\Omega}_p$ has unit length. From (2.2) and (2.6), the modulus of $F$ is

$$|F(\Omega, \Omega_p)| = \frac{1 - (\hat{\Omega} \cdot \hat{\Omega}_p)^2}{2(1 + \hat{\Omega} \cdot \hat{\Omega}_p)} = \frac{1}{2}(1 - \hat{\Omega} \cdot \hat{\Omega}_p). \quad (2.7)$$

Thus, the modulus of $F$ is completely determined by the angle between the GW direction and the pulsar direction. However, the phase of $F$, which depends upon the polarization of the GW, is not a function of this angle alone.

Since GWs at Earth are very weak, to obtain the instantaneous redshift at Earth for a given pulsar, we simply add up the real parts of $h(t)F^*$ for each source, where the real and imaginary parts of $h$ are the amplitudes at Earth of the two different polarizations. (If we were using timing residuals rather than redshift to describe the effect of GWs, then an integral over time would be needed rather than simply the instantaneous product.)

The antenna pattern function $F$ can be thought of as a propagator or response function which encodes the way that pulsar redshift responds to a GW. Physically, it is enough to specify this response for one source direction (say, $\hat{\Omega} = \hat{z}$) and all pulsar directions. Then, the response for any other source and pulsar directions can be obtained by rotation. This embodies a fundamental tenet of the principle of relativity: physical observables are coordinate-independent.

While this is true, there is an important subtlety. To explain it, let us start with the pulsar response for a gravitational wave propagating in the positive $z$ direction, obtained by setting $\theta = 0$ in (2.2). For that case $\hat{\Omega}_p \cdot (\hat{m} + i\hat{n}) = \sin\theta_p e^{i(\phi_p - \phi)}$, so by inspection one obtains

$$F(\hat{z}, \Omega_p) = \frac{1}{2}\frac{\sin^2\theta_p}{1 + \cos\theta_p}e^{2i(\phi_p - \phi)}$$
$$= \frac{1}{2}(1 - \cos\theta_p)e^{2i(\phi_p - \phi)}. \quad (2.8)$$

Note that this has a strange feature. Although $\theta = 0$ places the GW source direction at the North Pole for any value of $\phi$, the response still depends upon $\phi$. The reason has to do with the behavior of the polarization vectors $\hat{m}$ and $\hat{n}$. If we let the GW propagation direction $\hat{\Omega}$ approach the North Pole along different lines of longitude, the limiting values for the polarization vectors $\hat{m}$ and $\hat{n}$ depend upon which line of longitude is followed [41]. Hence, the response still depends upon $\phi$.

At the root of this odd behavior is the following observation. Since (2.8) gives the response of a pulsar at any point on the sky to a GW propagating in the $z$ direction, we should be able to determine the response of a pulsar in *any* direction, to a GW propagating in *any other* direction, simply by rotating the $z$ axis to the desired new propagation direction. But the rotation must not only carry the $\hat{z}$ vector to the new GW propagation direction: it must *also* carry the pair of vectors $\hat{m}$ and $\hat{n}$ to the correct ones at a different point on the sphere. To say it in another way, the response (2.2) only depends upon the dot products of different vectors, which are rotation-invariant. But, if the GW source is carried to a new sky position, then the corresponding vectors $\hat{m}$ and $\hat{n}$ must *also* be carried along in a way that matches their definitions in (2.3). If not, then $F$ rotates by a complex phase, so is not invariant. See the paragraph following (B5) for a precise statement.

There is a simple formula which encodes this complicated invariance in a beautiful way. If we first define a set of numerical coefficients by

$$A_l \equiv \frac{4\pi(-1)^l}{\sqrt{(l+2)(l+1)l(l-1)}}, \quad (2.9)$$

then

$$F(\Omega, \Omega_p) = \sum_{l=2}^{\infty}\sum_{m=-l}^{l} A_{l\;2}Y_{lm}(\Omega)Y^*_{lm}(\Omega_p). \quad (2.10)$$

This expression is extremely useful. In this paper, it plays a central role, similar to that of the spherical harmonic decomposition of the Green's function in electrostatics.

The relationship (2.10) is derived in Appendix B and is a mathematical equality: for any choice of the four arguments $\theta, \phi, \theta_p, \phi_p$, the right-hand sides of (2.2) and (2.10) yield the same complex number. Similar expressions can be found in the literature, for example, [14] [Eq. (18)] or [23] [Eq. (39)], though the latter has an undetermined complex phase.

For convenience, we define $A_0 = A_1 = 0$, so that sums like the one in (2.10) can be written $\sum_{lm}$. It is also helpful to define coefficients

$$a_l \equiv (2l+1)\left(\frac{A_l}{4\pi}\right)^2 = \begin{cases} 0 & \text{for } l < 2 \\ \frac{2l+1}{(l+2)(l+1)l(l-1)} & \text{for } l \geq 2. \end{cases} \quad (2.11)$$

These simplify the appearance of equations which follow.

The functions $Y_{lm}(\Omega_p)$ which appear on the rhs of (2.10) are the familiar spherical harmonics. These govern the way that the response varies with pulsar direction. In contrast, the $_2Y_{lm}(\Omega)$, through which the GW direction $\Omega$ enters the equation, are spin-2 weighted spherical harmonics. These spin-2 weighted harmonics form a complete orthonormal set on the unit sphere and have properties similar to normal spherical harmonics, for example, their $\phi$-dependence is $e^{im\phi}$. While only their general properties are needed for this paper, we give a precise definition in (A5), and full details may be found in [8] (Appendix A).

Note that for all spherical harmonics, we use the sign, phase and normalization conventions of [8] (Appendixes A and B), where a complete set of formulas is given; the most important ones are reproduced in Appendix A, and further details may be found in [42].

The representation of $F(\Omega, \Omega_p)$ given in (2.10) is a "factored" or "diagonal" form. In contrast, suppose that we tried to express $F$ in terms of ordinary spherical harmonics. Since it is a square integrable function of $\Omega$ and of $\Omega_p$, it can be decomposed as a sum of the form $F(\Omega, \Omega_p) = \sum_{lm} \sum_{l'm'} a_{lm,l'm'} Y_{lm}(\Omega) Y^*_{l'm'}(\Omega_p)$ for some set of expansion coefficients $a_{lm,l'm'}$. As discussed immediately after (2.3), $F(\Omega, \Omega_p)$ as defined by (2.2) only depends upon $\phi$ and $\phi_p$ through the difference $\phi - \phi_p$. Thus, the expansion coefficients $a_{lm,l'm'}$ vanish for $m \neq m'$. But, unlike the expansion in (2.10), the coefficients $a_{lm,l'm'}$ are *not* diagonal in $l$ and $l'$: $a_{lm,l'm} \neq 0$ for $l \neq l'$.

The individual plus- and cross-polarization components are easily extracted. Either from (2.2), or from (2.10) and (A3), one can immediately see that $F^*(\Omega, \Omega_p) = F(\bar{\Omega}, \bar{\Omega}_p)$. Here, the overlines indicate antipodal points on the sphere: $\bar{\Omega}$ has coordinates $\bar{\theta} = \pi - \theta$, $\bar{\phi} = \phi + \pi$, and $\bar{\Omega}_p$ has coordinates $\bar{\theta}_p = \pi - \theta_p$, $\bar{\phi}_p = \phi_p + \pi$. Thus,

$$F^+(\Omega, \Omega_p) = \frac{1}{2}\left(F(\Omega, \Omega_p) + F(\bar{\Omega}, \bar{\Omega}_p)\right),$$
$$F^\times(\Omega, \Omega_p) = \frac{1}{2i}\left(F(\Omega, \Omega_p) - F(\bar{\Omega}, \bar{\Omega}_p)\right). \quad (2.12)$$

This is also how the individual polarization components were extracted in [27].

The correlation between pulsars $p$ and $q$ is a function $\varrho_{pq}(\Omega)$ of their directions, and of the propagation direction $\Omega$ of the GW source. This is often called the "HD integrand" and can be written in several equivalent forms:

$$\varrho_{pq}(\Omega) \equiv \Re[F(\Omega, \Omega_p)F^*(\Omega, \Omega_q)]$$
$$= \frac{1}{2}[F(\Omega, \Omega_p)F^*(\Omega, \Omega_q) + F(\Omega, \Omega_q)F^*(\Omega, \Omega_p)]$$
$$= \frac{1}{2}[F(\Omega, \Omega_p)F^*(\Omega, \Omega_q) + F(\bar{\Omega}, \bar{\Omega}_p)F^*(\bar{\Omega}, \bar{\Omega}_q)]$$
$$= F^+(\Omega, \Omega_p)F^+(\Omega, \Omega_q) + F^\times(\Omega, \Omega_p)F^\times(\Omega, \Omega_q). \quad (2.13)$$

The second equality shows that the real part of $FF^*$ may be obtained from $FF^*$ by swapping the locations of the two pulsars.

## III. THE HELLINGS AND DOWNS CURVE AS AN AVERAGE OVER SOURCE DIRECTIONS

The HD curve $\mu_u(\gamma)$ was originally defined [26] as the correlation between two pulsars $p$ and $q$ separated by angle $\gamma$ in the sky, uniformly averaged over source directions, for a unit amplitude unpolarized source. We use

$$\int d\Omega \equiv \int_0^\pi \sin\theta \, d\theta \int_0^{2\pi} d\phi \quad (3.1)$$

to denote the integral over the unit two-sphere. To average over directions, an additional factor of $1/4\pi$ must be included. The angle between the pulsars is

$$\cos\gamma = \cos\gamma_{pq} \equiv \hat{\Omega}_p \cdot \hat{\Omega}_q$$
$$= \cos\theta_p \cos\theta_q + \sin\theta_p \sin\theta_q \cos(\phi_p - \phi_q), \quad (3.2)$$

where the final equality follows immediately from (2.1).

The computation of the HD curve, starting from (2.10), is trivial. We denote the sky locations of the two pulsars by $\Omega_p$ and $\Omega_q$, and let $\varrho_{pq}(\Omega)$ given in (2.13) denote their correlation. The average of this is

$$\mu_u = \frac{1}{4\pi}\int d\Omega \, \varrho_{pq}(\Omega)$$
$$= \frac{1}{4\pi}\Re\int d\Omega \, F(\Omega, \Omega_p)F^*(\Omega, \Omega_q)$$
$$= \frac{1}{4\pi}\Re\sum_{lm}\sum_{l'm'} A_l A_{l'} Y^*_{lm}(\Omega_p)Y_{l'm'}(\Omega_q)$$
$$\times \int d\Omega \, _2Y_{lm}(\Omega) \, _2Y^*_{l'm'}(\Omega)$$
$$= \frac{1}{4\pi}\Re\sum_{lm}\sum_{l'm'} A_l A_{l'} Y^*_{lm}(\Omega_p)Y_{l'm'}(\Omega_q)\delta_{ll'}\delta_{mm'}$$
$$= \frac{1}{4\pi}\Re\sum_{lm} A_l^2 Y^*_{lm}(\Omega_p)Y_{lm}(\Omega_q)$$
$$= \frac{1}{4\pi}\sum_l A_l^2 \left(\frac{2l+1}{4\pi}\right) P_l(\hat{\Omega}_p \cdot \hat{\Omega}_q)$$
$$= \sum_l a_l P_l(\cos\gamma), \quad (3.3)$$

where $P_l(z)$ is the Legendre polynomial of order $l$, and $\gamma = \cos^{-1}(\hat{\Omega}_p \cdot \hat{\Omega}_q)$ is the angle between the lines of sight to the two pulsars. The second equality follows directly from the definition (2.13), the third from (2.10), the fourth and fifth equalities follow because the spin-2 weighted harmonics form an orthonormal set on the unit sphere, the sixth equality follows from the addition theorem for spherical harmonics,

$$P_l(\hat{\Omega}_p \cdot \hat{\Omega}_q) = \frac{4\pi}{2l+1} \sum_{m=-l}^{l} Y_{lm}(\Omega_p) Y_{lm}^*(\Omega_q), \quad (3.4)$$

and the final equality follows from the definitions of $A_l$ and $a_l$ in (2.9) and (2.11).

The final expression in (3.3) is the standard harmonic-space form of the famous HD curve $\mu_u(\gamma)$. One can easily carry out the sum [8] (Sec. III.E) to obtain the position space form

$$\mu_u(\gamma) = \frac{1}{3} + \frac{1}{2}(1 - \cos\gamma)\left[-\frac{1}{6} + \log\left(\frac{1-\cos\gamma}{2}\right)\right]. \quad (3.5)$$

Note that the same result would have been obtained without taking the real part on the second line of (3.3). This is because the imaginary part of $F(\Omega,\Omega_p)F^*(\Omega,\Omega_q)$ is odd under $\Omega \to \bar{\Omega}$, so it integrates to zero.

To reduce clutter, we often omit the summation limits on $l$ and $m$. In such cases, $l = 0, 1, 2, \ldots$ and $m = -l, -l+1, \ldots, l-1, l$. Here, the sum is effectively over $l = 2, 3, \ldots$, because $a_l$ and $A_l$ vanish for $l < 2$.

## IV. PULSAR AVERAGING

"Pulsar averaging" is a useful calculational method, which was first introduced in [28] and then developed further in [27,38]. It is defined as follows. Given a function $Q(\Omega_p, \Omega_q)$ which depends upon the position of two pulsars $p$ and $q$, the pulsar average of $Q$ is a function of angle $\gamma \in [0, \pi]$, and is defined by

$$\begin{aligned}Q(\gamma) &= \langle Q(\Omega_p, \Omega_q) \rangle_{pq \in \gamma} \\ &\equiv \frac{1}{8\pi^2} \int d\Omega_p \int d\Omega_q \, \delta(\hat{\Omega}_p \cdot \hat{\Omega}_q - \cos\gamma) \\ &\quad \times Q(\Omega_p, \Omega_q),\end{aligned} \quad (4.1)$$

where $\delta(x)$ is the ordinary Dirac delta function. Replacing $Q$ by the constant function $Q = 1$, one can easily verify that this average is correctly normalized, meaning that $\langle 1 \rangle_{pq \in \gamma} = 1$.

This definition corresponds to an average over all unit vectors $\hat{\Omega}_p$ uniformly distributed on the sphere, and all unit vectors $\hat{\Omega}_q$ uniformly distributed in a cone at angle $\gamma$ around $\hat{\Omega}_p$, as illustrated in Fig. 5 of [27]. It has a close analog in experimental practice, for example, when the Hellings and Downs curve is "reconstructed" by binning together measured correlations from large numbers of pulsar pairs [38] with similar separation angles.

For calculational purposes, it is helpful to express the Dirac delta function in (4.1) in terms of spherical harmonics. To do this, begin with the Dirac delta function expressed as a sum of Legendre polynomials $P_l(x)$, as derived in Eq. (4.20) of [38]. On the interval $x, x' \in [-1, 1]$,

$$\delta(x - x') = \sum_l \frac{2l+1}{2} P_l(x) P_l(x'). \quad (4.2)$$

In (4.2), set $x = \hat{\Omega}_p \cdot \hat{\Omega}_q$ and $x' = \cos\gamma$ on the lhs, and on the rhs replace $P_l(\hat{\Omega}_p \cdot \hat{\Omega}_q)$ using the addition theorem (3.4). This gives

$$\delta(\hat{\Omega}_p \cdot \hat{\Omega}_q - \cos\gamma) = 2\pi \sum_{lm} P_l(\cos\gamma) Y_{lm}(\Omega_p) Y_{lm}^*(\Omega_q). \quad (4.3)$$

Note that setting $\gamma = 0$ correctly implies that $\delta^2(\Omega_p, \Omega_q) = (1/2\pi)\delta(\hat{\Omega}_p \cdot \hat{\Omega}_q - 1)$, where the lhs is the two-dimensional delta function on the unit two-sphere $S^2$.

If we return to the definition (4.1) of the pulsar average, and replace the delta function with (4.3), we obtain

$$\begin{aligned}Q(\gamma) &= \langle Q(\Omega_p, \Omega_q) \rangle_{pq \in \gamma} \\ &= \frac{1}{4\pi} \int d\Omega_p \int d\Omega_q \sum_{lm} P_l(\cos\gamma) Y_{lm}(\Omega_p) Y_{lm}^*(\Omega_q) \\ &\quad \times Q(\Omega_p, \Omega_q).\end{aligned} \quad (4.4)$$

This recipe for computing the pulsar average of any function $Q(\Omega_p, \Omega_q)$ of pulsar positions will be used later for computing the total variance in models with correlated GW source sky locations.

The pulsar average of the function $Q(\Omega_p, \Omega_q) = Y_{lm}(\Omega_p) Y_{l'm'}^*(\Omega_q)$ will be needed later. This is evaluated starting from the definition (4.1) as

$$\begin{aligned}&\langle Y_{lm}(\Omega_p) Y_{l'm'}^*(\Omega_q) \rangle_{pq \in \gamma} \\ &\equiv \frac{1}{8\pi^2} \int d\Omega_p \int d\Omega_q \, \delta(\hat{\Omega}_p \cdot \hat{\Omega}_q - \cos\gamma) Y_{lm}(\Omega_p) Y_{l'm'}^*(\Omega_q) \\ &= \frac{1}{4\pi} \sum_{l''m''} P_{l''}(\cos\gamma) \int d\Omega_p Y_{lm}(\Omega_p) Y_{l''m''}^*(\Omega_p) \times \\ &\qquad\qquad \int d\Omega_q Y_{l'm'}^*(\Omega_q) Y_{l''m''}(\Omega_q) \\ &= \frac{1}{4\pi} \sum_{l''m''} P_{l''}(\cos\gamma) \delta_{ll''} \delta_{mm''} \delta_{l'l''} \delta_{m'm''} \\ &= \frac{1}{4\pi} \delta_{ll'} \delta_{mm'} P_l(\cos\gamma).\end{aligned} \quad (4.5)$$

Here, the second equality is obtained using (4.3), the third equality follows from the orthonormality of the spherical harmonics, and the final equality from the definition of the Kronecker delta. We now use this to carry out some additional harmonic-space computations.

## V. THE HELLINGS AND DOWNS CURVE AS A PULSAR AVERAGE FOR A SINGLE GW SOURCE

An alternative definition of the HD curve is as the pulsar average of the cross-correlation (2.13) for one fixed circularly polarized GW point source. This approach was first investigated in [28] and then further developed in [27]. Here, we compute this pulsar average, starting from the harmonic expansion (2.10) of the response function $F$.

For this computation, we fix $\Omega$, and compute the pulsar average of the correlation (2.13)

$$\begin{aligned}
\langle \varrho_{pq}(\Omega) \rangle_{pq \in \gamma} &= \Re \langle F(\Omega, \Omega_p) F^*(\Omega, \Omega_q) \rangle_{pq \in \gamma} \\
&= \Re \sum_{lm} \sum_{l'm'} A_l A_{l'} \, {}_2Y_{lm}(\Omega) \, {}_2Y^*_{l'm'}(\Omega) \\
&\quad \times \langle Y^*_{lm}(\Omega_p) Y_{l'm'}(\Omega_q) \rangle_{pq \in \gamma} \\
&= \frac{1}{4\pi} \Re \sum_l A_l^2 P_l(\cos\gamma) \sum_{m=-l}^{l} {}_2Y_{lm}(\Omega) \, {}_2Y^*_{lm}(\Omega) \\
&= \frac{1}{4\pi} \sum_l \left( \frac{2l+1}{4\pi} \right) A_l^2 P_l(\cos\gamma) \\
&= \sum_l a_l P_l(\cos\gamma) \\
&= \mu_{\mathrm{u}}(\gamma).
\end{aligned} \quad (5.1)$$

The second equality is obtained by substituting the diagonal form (2.10) for $F$, the third by substituting the pulsar average of two spherical harmonics given by (4.5), the fourth from the sum of spin-2 weighted harmonics

$$\sum_m {}_2Y_{lm}(\Omega) \, {}_2Y^*_{lm}(\Omega) = \frac{2l+1}{4\pi}, \quad (5.2)$$

the fifth equality follows from the definition of $a_l$ in (2.11) and the final equality from comparison with the average over source directions computed in (3.3). This equality, between (a) the pulsar average for a single source and (b) the average response of a single pair of pulsars to an isotropically distributed set of (noninterfering) sources, was first demonstrated in [28].

In the next section, we will discuss the addition theorem for spin-2 weighted spherical harmonics, from which (5.2) may be obtained as a special case by setting $\beta = \chi = 0$ in (6.4).

## VI. THE HELLINGS AND DOWNS TWO-POINT FUNCTION

Previous work [27,38,43,44] on the variance of the HD correlation exploited a two-point function. Here, this is defined in analogy with Eq. (G1) of [27] as

$$\mu(\gamma, \Omega, \Omega') \equiv \langle F(\Omega, \Omega_p) F^*(\Omega', \Omega_q) \rangle_{pq \in \gamma}. \quad (6.1)$$

This is averaging the complex redshift response of a pulsar with sky direction $\hat{\Omega}_p$ to a distant unit-amplitude GW point source with sky direction $-\hat{\Omega}$, with the corresponding response for a second pulsar $\hat{\Omega}_q$ to a second unit-amplitude point source with sky direction $-\hat{\Omega}'$. [The minus signs are explained after (2.1).] As before, $\gamma$ is the angular separation on the sky of the two pulsars.

The original definition given in [27] is slightly different: it is a real quantity $\mu(\gamma, \beta)$ whose square is the squared modulus of the complex quantity $\mu(\gamma, \Omega, \Omega')$ defined here. We will see that the modulus depends upon the directions to the two GW point sources only via the angle $\beta \in [0, \pi]$ between their lines of sight, where

$$\cos\beta \equiv \hat{\Omega} \cdot \hat{\Omega}' = \cos\theta \cos\theta' + \sin\theta \sin\theta' \cos(\phi - \phi'). \quad (6.2)$$

The magnitude $\mu^2(\gamma, \beta) = |\mu(\gamma, \Omega, \Omega')|^2$ is what matters: the phase of $\mu(\gamma, \Omega, \Omega')$ is a "gauge artifact" that drops out of observable quantities.

To evaluate the two-point function (6.1), we substitute $F$ from (2.10) into the definition and use (4.5) to compute the pulsar average of $Y_{lm}(\Omega_p) Y^*_{l'm'}(\Omega_q)$. We obtain

$$\begin{aligned}
\mu(\gamma, \Omega, \Omega') &= \sum_{lm} \sum_{l'm'} A_l A_{l'} \, {}_2Y_{lm}(\Omega) \, {}_2Y^*_{l'm'}(\Omega') \\
&\quad \times \langle Y_{lm}(\Omega_p) Y^*_{l'm'}(\Omega_q) \rangle_{pq \in \gamma} \\
&= \frac{1}{4\pi} \sum_{lm} A_l^2 P_l(\cos\gamma) \, {}_2Y_{lm}(\Omega) \, {}_2Y^*_{lm}(\Omega') \\
&= \frac{1}{4\pi} \sum_l A_l^2 P_l(\cos\gamma) \sum_{m=-l}^{l} {}_2Y_{lm}(\Omega) \, {}_2Y^*_{lm}(\Omega').
\end{aligned} \quad (6.3)$$

The final sum over $m$ is the spin-2 equivalent of the traditional addition theorem (3.4) for scalar harmonics.

The addition theorem for spin-weighted harmonics is given in [8] [(A9)–(A11)]. Using [8] [(A6)] and the relation between the Wigner "big D" and "small d" matrices $D^j_{m'm}(\phi, \theta, \psi) = e^{-im'\phi} d^j_{m'm}(\theta) e^{-im\psi}$, the sum appearing in (6.3) may be written

$$\sum_{m=-l}^{l} {}_2Y_{lm}(\Omega) \, {}_2Y^*_{lm}(\Omega')$$
$$= \frac{2l+1}{4\pi} \left( \cos\frac{\beta}{2} \right)^4 P^{(0,4)}_{l-2}(\cos\beta) e^{2i\chi(\Omega,\Omega')}, \quad (6.4)$$

where we have expressed the Wigner small d matrix in terms of Jacobi polynomials. These are polynomials in $\sin^2(\beta/2)$ and $\cos^2(\beta/2)$, and are illustrated in Fig. 1.

In contrast with the corresponding sum of scalar harmonics, the sum on the final line of (6.3) does not just depend upon the angular separation $\beta$ between $\Omega$ and $\Omega'$.

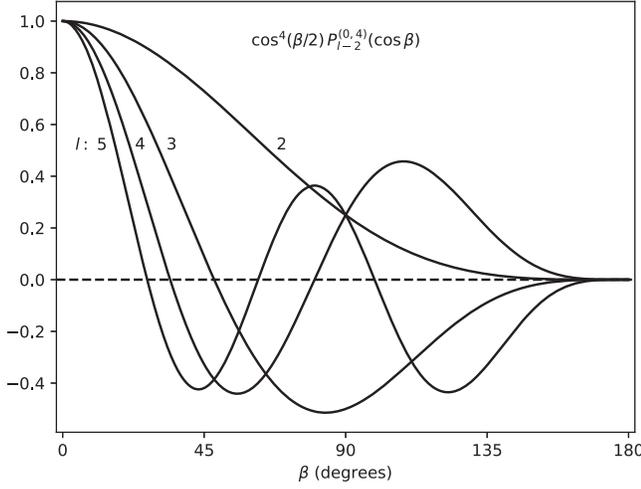

FIG. 1. The Jacobi polynomials of (6.4) are shown for $l = 2, \ldots, 5$.

While the magnitude of (6.4) is only a function of $\beta$, its phase has a complicated dependence upon the positions of the two GW sources. This dependence is via the real angle $\chi$ defined by

$$\tan\frac{1}{2}\chi(\Omega,\Omega') \equiv \frac{\sin\frac{1}{2}(\phi'-\phi)\cos\frac{1}{2}(\theta+\theta')}{\cos\frac{1}{2}(\phi'-\phi)\cos\frac{1}{2}(\theta'-\theta)}. \quad (6.5)$$

In the notation of [8], $\chi = \phi_3 + \chi_3$.

The angle $\chi(\Omega,\Omega') \in [-\pi,\pi]$ may be defined by inverting (6.5), with arctan in the range $[-\pi/2, \pi/2]$ or in the range $[0, \pi]$. Alternatively, $\chi$ may be defined in the range $[0, 4\pi]$ as the argument of the complex number whose imaginary and real parts are (respectively) the numerator and denominator in (6.5). Because $\chi$ only enters (6.4) via $e^{2i\chi}$, these different choices are equivalent.

An important property of $\chi$ is that it is an antisymmetric function of its two arguments:

$$\chi(\Omega,\Omega') = -\chi(\Omega',\Omega). \quad (6.6)$$

This proves that $\chi$ cannot be written as a function of $\beta$, since $\beta$ is a symmetric function of $\Omega$ and $\Omega'$. Another important property of $\chi$, which also follows directly from its definition (6.5), is that $\chi$ changes sign if both arguments are sent to their antipodal points. Using the notation introduced in (2.12), this is written

$$\chi(\bar{\Omega},\bar{\Omega}') = -\chi(\Omega,\Omega'). \quad (6.7)$$

But $\chi$ is mostly a nuisance: as discussed in [27] (Appendix G), we will see that $e^{2i\chi}$ is a gauge artifact that drops out of physically observable quantities.

Making use of the addition theorem for spin-weighted spherical harmonics provides an elegant harmonic decomposition of the two-point function. Substituting the sum over $m$ in (6.4) into (6.3), and using $a_l$ from (2.11) gives

$$\mu(\gamma,\Omega,\Omega') = \mu(\gamma,\beta)e^{2i\chi(\Omega,\Omega')}, \quad (6.8)$$

where

$$\mu(\gamma,\beta) = \left(\cos\frac{\beta}{2}\right)^4 \sum_l a_l P_{l-2}^{(0,4)}(\cos\beta)P_l(\cos\gamma). \quad (6.9)$$

Note that in these equations, the real quantity $\mu(\gamma,\beta)$ may have either sign, so it cannot be interpreted as a radius in the complex plane.

As it must, the two-point function (6.8) reduces to the normal HD curve in the limit of coincident GW sources $\Omega' \to \Omega$, where $\beta \to 0$ and $\chi \to 0$. Since the Jacobi polynomials are normalized to $P_l^{(\alpha,\beta)}(1) = 1$, (6.8) and (6.9) immediately give $\mu(\gamma,\Omega,\Omega) = \mu(\gamma,0) = \mu_u(\gamma)$, in agreement with the HD curve of (3.3).

## VII. COSMIC VARIANCE AND COVARIANCE AND GAUGE INDEPENDENCE

We now investigate the complex phase $e^{2i\chi(\Omega,\Omega')}$ which appears in (6.8), and show that it drops out of the cosmic covariance, which is a physical observable. This also establishes the gauge independence of the cosmic variance, since it is the covariance restricted to the diagonal. While the results are more general, here we demonstrate them for the specific case of the Gaussian ensemble [38].

The GW metric perturbations in any representative universe may be defined via a plane-wave expansion [27] [Eq. (C1)]

$$h_{ab}(t,\vec{x}) = \Re\int df\int d\Omega\, h(f,\Omega)e^*_{ab}(\Omega)e^{2\pi i f(t-\hat{\Omega}\cdot\vec{x})}, \quad (7.1)$$

where we use a complex polarization basis to simplify what follows. Each realization of the universe is defined by its own specific complex Fourier amplitudes $h(f,\Omega)$. In terms of the linear polarization basis of [27] (Appendix C), these are

$$h(f,\Omega) = h_+(f,\Omega) + ih_\times(f,\Omega), \quad (7.2)$$

$$e_{ab}(\Omega) = e^+_{ab}(\Omega) + ie^\times_{ab}(\Omega), \quad (7.3)$$

where the reader should keep in mind that $h_+(f,\Omega) = h^*_+(-f,\Omega)$ and $h_\times(f,\Omega) = h^*_\times(-f,\Omega)$ are complex quantities. An ensemble is defined by a set of Fourier amplitude functions $h(f,\Omega)$. Each specific function corresponds to a particular universe within the ensemble.

An ensemble may equivalently be defined by specifying all moments of $h(f,\Omega)$. Letting angle brackets $\langle\rangle$ denote averages over that ensemble, the Gaussian ensemble is fully defined by the first and second moments

$$\langle h(f, \Omega) \rangle = 0,$$
$$\langle h(f, \Omega) h(f', \Omega') \rangle = 0,$$
$$\langle h(f, \Omega) h^*(f', \Omega') \rangle = 2H(f)\delta(f - f')\delta^2(\Omega, \Omega'), \quad (7.4)$$

taken together with Isserlis's theorem [45]. Here, $H(f) = H(-f)$ is a real spectral function, and the factor of two is to maintain notational consistency with [27] [Eq. (C4)] and [38]. Isserlis's theorem defines the higher-order moments $\langle h(f, \Omega) h(f', \Omega')...h(f'', \Omega'') \rangle$, where any of the functions might also be complex-conjugated, in terms of the first and second moments given by (7.4).

The relations in (7.4) are usually stated for plus- and cross-polarization components. At first glance it appears that (7.4) provides two second moments, whereas the conventional expressions have only one. That is misleading: the conventional expressions have four second moments, for the four combinations of plus and cross, two of which vanish. If we had used a right- and left-circular polarization basis, then the last two lines of (7.4) could be combined into a single equation, with a Kronecker delta for the two polarization states on the rhs. The apparent extra factor of two arises because the $\langle hh^* \rangle$ term is the sum of the linear polarization plus-plus and cross-cross terms.

We define the correlation $\rho_{pq}$ between pulsars $p$ and $q$ following Eq. (C15) of [27]. For any representative universe in the ensemble, the correlation between pulsars is

$$\rho_{pq} \equiv \overline{Z_p(t) Z_q(t)}, \quad (7.5)$$

where $Z_p(t)$ is the (real, physical) redshift of pulsar $p$ as a function of time, and overline denotes a time average. In what follows, the averaging-time interval is denoted $T$, which may equivalently be taken as the total observation time.

The pulsar-averaged correlation $\Gamma(\gamma)$ is defined following Eq. (C41) of [27], as

$$\Gamma(\gamma) \equiv \langle \rho_{pq} \rangle_{pq \in \gamma}. \quad (7.6)$$

Here, the angle brackets denote the average over all pulsar pairs $p$ and $q$ separated by angle $\gamma$ on the sky, as defined in Sec. IV.

The pulsar-averaged correlation in any representative universe in the ensemble may be computed in the same way as [27] [Eq. (C41)]. It is

$$\Gamma(\gamma) = \frac{1}{4} \int df \int df' \int d\Omega \int d\Omega' \text{sinc}(\pi(f - f')T) \times$$
$$\left\langle \left[ h^*(f', \Omega') F(\Omega', \Omega_q) + h(f', \Omega') F^*(\Omega', \Omega_q) \right] \times \right.$$
$$\left. \left[ h(f, \Omega) F^*(\Omega, \Omega_p) + h^*(f, \Omega) F(\Omega, \Omega_p) \right] \right\rangle_{pq \in \gamma}$$
$$= \frac{1}{4} \int df \int df' \int d\Omega \int d\Omega' \text{sinc}(\pi(f - f')T) \times$$
$$\left[ h(f, \Omega) \ h^*(f', \Omega') \mu(\gamma, \bar{\Omega}, \bar{\Omega}') + \right.$$
$$h(f, \Omega) \ h(f', \Omega') \ \mu(\bar{\gamma}, \bar{\Omega}, \Omega') +$$
$$h^*(f, \Omega) h^*(f', \Omega') \mu(\bar{\gamma}, \Omega, \bar{\Omega}') +$$
$$\left. h^*(f, \Omega) h(f', \Omega') \ \mu(\gamma, \Omega, \Omega') \right], \quad (7.7)$$

where $\bar{\gamma} \equiv \pi - \gamma$ is the angular sky separation between $\Omega_p$ and $\bar{\Omega}_q$ or between $\Omega_q$ and $\bar{\Omega}_p$, and $\text{sinc} x \equiv \sin(x)/x$. The first equality follows from the definition of $\Gamma(\gamma)$ as the pulsar-averaged correlation at angle $\gamma$, with the factors of $1/2$ arising from taking the real part as given in (7.1). (As shown in [27], only Earth terms survive the pulsar averaging, so pulsar terms have been dropped.) The second equality follows from the definition (6.1) of the two-point function $\mu(\gamma, \Omega, \Omega')$ and the use of $F^*(\Omega, \Omega_p) = F(\bar{\Omega}, \bar{\Omega}_p)$ to generate complex conjugates of $F$, as previously employed in (2.12).

The ensemble average of $\Gamma$ may be computed from inspection of (7.7), using the second moments (7.4) for the Gaussian ensemble. The second and third terms vanish, and the first and fourth terms give

$$\langle \Gamma(\gamma) \rangle = \frac{1}{2} \int H(f) df \int \left( \mu(\gamma, \bar{\Omega}, \bar{\Omega}) + \mu(\gamma, \Omega, \Omega) \right) d\Omega$$
$$= h^2 \mu_u(\gamma). \quad (7.8)$$

To obtain the final equality, we have used the fact that $\mu(\gamma, \Omega, \Omega) = \mu_u(\gamma)$ is the HD curve, and independent of source direction $\Omega$. The squared GW strain at Earth

$$h^2 \equiv 4\pi \int H(f) df \quad (7.9)$$

is defined using notation compatible with [27] and [38].

To compute the covariance and variance, we need the deviation of the correlation away from the mean, for any representative of the ensemble. This is

$$\Delta \Gamma(\gamma) \equiv \Gamma(\gamma) - \langle \Gamma(\gamma) \rangle, \quad (7.10)$$

where, as before, angle brackets with no trailing subscript denote an ensemble average. It follows immediately from the definition above that $\langle \Delta\Gamma(\gamma) \rangle$ vanishes. The cosmic covariance is the ensemble average

$$\sigma^2_{\rm cos}(\gamma,\gamma') \equiv \langle \Delta\Gamma(\gamma)\Delta\Gamma(\gamma')\rangle$$
$$= \langle \Gamma(\gamma)\Gamma(\gamma')\rangle - \langle \Gamma(\gamma)\rangle\langle\Gamma(\gamma')\rangle. \quad (7.11)$$

Note that the cosmic covariance may have either sign, whereas the cosmic variance (the value of the covariance along the diagonal $\gamma = \gamma'$) is non-negative. Notationally, they are easily distinguished, because the cosmic variance has one argument, whereas the cosmic covariance has two.

The cosmic covariance $\sigma^2_{\rm cos}(\gamma,\gamma')$ can be computed directly from (7.7). The expression for $\Gamma(\gamma)\Gamma(\gamma')$ contains 16 terms, whose average over the Gaussian ensemble can be evaluated using Isserlis's theorem. This implies that

$$\langle h(f,\Omega)h^*(f',\Omega')h(f'',\Omega'')h^*(f''',\Omega''')\rangle$$
$$= \langle h(f,\Omega)h^*(f',\Omega')\rangle\langle h(f'',\Omega'')h^*(f''',\Omega''')\rangle$$
$$+ \langle h(f,\Omega)h^*(f''',\Omega''')\rangle\langle h(f'',\Omega'')h^*(f',\Omega')\rangle, \quad (7.12)$$

and that the ensemble average of terms with unequal numbers of $h$ and $h^*$ vanish.

To evaluate the cosmic covariance $\sigma^2_{\rm cos}(\gamma,\gamma')$, we start by noting that among the 16 terms of $\Gamma(\gamma)\Gamma(\gamma')$ are 10 terms containing unequal numbers of $h$ and $h^*$; their ensemble averages vanish. Each of the remaining six terms contains two delta functions in frequency and two delta functions on the sphere. Integrating those out gives

$$\langle\Gamma(\gamma)\Gamma(\gamma')\rangle = \langle\Gamma(\gamma)\rangle\langle\Gamma(\gamma')\rangle + \frac{\hbar^4}{4}\int\frac{d\Omega}{4\pi}\int\frac{d\Omega'}{4\pi}\times$$
$$\Big[\mu(\gamma,\Omega,\Omega')\big(\mu(\gamma',\Omega',\Omega)+\mu(\gamma',\bar\Omega,\bar\Omega')\big)+$$
$$\mu(\gamma,\bar\Omega,\bar\Omega')\big(\mu(\gamma',\Omega,\Omega')+\mu(\gamma',\bar\Omega',\bar\Omega)\big)+$$
$$\mu(\bar\gamma,\bar\Omega,\Omega')\big(\mu(\bar\gamma',\Omega,\bar\Omega')+\mu(\bar\gamma',\Omega',\bar\Omega)\big)+$$
$$\mu(\bar\gamma,\Omega,\bar\Omega')\big(\mu(\bar\gamma',\bar\Omega,\Omega')+\mu(\bar\gamma',\bar\Omega',\Omega)\big)\Big], \quad (7.13)$$

where $\bar\gamma' \equiv \pi - \gamma'$ and we have defined (see Appendixes A and B of [38])

$$\hbar^4 \equiv (4\pi)^2\int df\int df' H(f)H(f'){\rm sinc}^2\big(\pi(f-f')T\big). \quad (7.14)$$

The key point is that the complex phase $\exp(2i\chi)$ cancels out of the cosmic covariance, as can be seen by inspection of (7.13). For example, the first two terms are

$$\mu(\gamma,\Omega,\Omega')\big(\mu(\gamma',\Omega',\Omega)+\mu(\gamma',\bar\Omega,\bar\Omega')\big)$$
$$= \mu(\gamma,\beta)e^{2i\chi(\Omega,\Omega')}\big(\mu(\gamma',\beta)e^{2i\chi(\Omega',\Omega)}+\mu(\gamma',\beta)e^{2i\chi(\bar\Omega,\bar\Omega')}\big)$$
$$= 2\mu(\gamma,\beta)\mu(\gamma',\beta). \quad (7.15)$$

The first equality is obtained using the harmonic form (6.8) for the two-point function, and the second equality follows from the antisymmetry of $\chi$ under interchange of the arguments (6.6) or antipodal reflection (6.7). Carrying out similar phase cancellations for the remaining terms in (7.13) yields a simple expression for the cosmic covariance. If we let

$$\sigma^2_{\rm cos}(\gamma,\gamma') = 2\hbar^4\widetilde{\mu^2}(\gamma,\gamma'), \quad (7.16)$$

then the dimensionless function that describes the cosmic covariance is

$$\widetilde{\mu^2}(\gamma,\gamma') \equiv \frac{1}{2}\int\frac{d\Omega}{4\pi}\int\frac{d\Omega'}{4\pi}\Big(\mu(\gamma,\beta)\mu(\gamma',\beta)+\mu(\bar\gamma,\bar\beta)\mu(\bar\gamma',\bar\beta)\Big)$$
$$= \frac{1}{4}\int_0^\pi d\beta\sin\beta\Big(\mu(\gamma,\beta)\mu(\gamma',\beta)+\mu(\bar\gamma,\beta)\mu(\bar\gamma',\beta)\Big). \quad (7.17)$$

The first equality follows from (7.11), (7.13), and simplifications such as those in (7.15). The second equality holds because the integral over the sphere is not changed if carried out with respect to the antipodal point (i.e., it is invariant under $\Omega' \to \bar\Omega'$ which is $\beta \to \bar\beta$).

The cosmic variance is obtained from the covariance by sending $\gamma' \to \gamma$, giving

$$\sigma^2_{\rm cos}(\gamma) \equiv \langle\Delta\Gamma(\gamma)^2\rangle$$
$$= \sigma^2_{\rm cos}(\gamma,\gamma)$$
$$= \frac{1}{2}\hbar^4\int_0^\pi d\beta\sin\beta\Big(\mu^2(\gamma,\beta)+\mu^2(\pi-\gamma,\beta)\Big)$$
$$= 2\hbar^4\widetilde{\mu^2}(\gamma), \quad (7.18)$$

which should be compared with [27] [Eq. (G12)] and is identical to [38] [Eq. (4.32)].

## VIII. HARMONIC FORM OF THE COSMIC VARIANCE AND COVARIANCE

Starting from the harmonic decomposition (6.9) of the two-point function, it is straightforward to obtain a harmonic form for the cosmic variance and covariance.

The Jacobi polynomials satisfy the orthogonality condition

$$\int_{-1}^1 dz(1-z)^a(1+z)^b P_l^{(a,b)}(z)P_{l'}^{(a,b)}(z)$$
$$= \frac{2^{a+b+1}(l+a)!(l+b)!}{(2l+a+b+1)l!(l+a+b)!}\delta_{ll'}, \quad (8.1)$$

where it is assumed that $l+1$, $a+1$ and $b+1$ are positive integers. For the case of interest, set $a=0$, $b=4$, and define $z = \cos\beta$. Since $1+z = 2\cos^2(\beta/2)$, if we send

$l \to l-2$ and $l' \to l'-2$ in (8.1), the orthogonality condition takes the form

$$\int_0^\pi \sin\beta \, d\beta \left(\cos\frac{\beta}{2}\right)^8 P_{l-2}^{(0,4)}(\cos\beta) P_{l'-2}^{(0,4)}(\cos\beta) = \frac{2\delta_{ll'}}{2l+1}. \quad (8.2)$$

This allows the integrals appearing in (7.17) to be evaluated by inspection.

The harmonic form of the cosmic covariance is obtained from (7.17) by replacing the two-point functions with the harmonic sums given in (6.9), and integrating using (8.2). The first of the two integrals is

$$\int_0^\pi d\beta \, \sin\beta \mu(\gamma,\beta)\mu(\gamma',\beta)$$
$$= \sum_{l,l'=2}^\infty \frac{2}{2l+1} \delta_{ll'} a_l a_{l'} P_l(\cos\gamma) P_{l'}(\cos\gamma')$$
$$= \sum_l \frac{2}{2l+1} a_l^2 P_l(\cos\gamma) P_l(\cos\gamma'), \quad (8.3)$$

where the coefficients $a_l$ are positive quantities defined in (2.11). With (8.3), it is easy to see that the second of the two integrals in (7.17) is equal to the first, since $P_l(\cos\bar{\gamma}) = (-1)^l P_l(\cos\gamma)$. Hence, from (7.17) and (8.3) we obtain a beautiful harmonic expansion of the cosmic covariance

$$\widetilde{\mu^2}(\gamma,\gamma') = \sum_l \frac{a_l^2}{2l+1} P_l(\cos\gamma) P_l(\cos\gamma'). \quad (8.4)$$

In the same way, we can evaluate cosmic variance function $\widetilde{\mu^2}(\gamma)$ defined by (7.18). This function encodes the angular ($\gamma$) dependence of the cosmic variance $\sigma_{\cos}^2(\gamma)$ for the Gaussian ensemble. From (7.18) it is

$$\widetilde{\mu^2}(\gamma) = \widetilde{\mu^2}(\gamma,\gamma) = \sum_l \frac{a_l^2}{2l+1} P_l^2(\cos\gamma), \quad (8.5)$$

where $P_l^2$ denotes the square of a Legendre polynomial and *not* the associated Legendre function with $m=2$. This harmonic form of the cosmic variance was first given in [27] [Eq. (C53)] and was found independently in [10].

## IX. MEAN AND VARIANCE OF HD CORRELATION IN MODELS WITH CORRELATED SOURCE SKY LOCATIONS

On the large scale, the universe appears to be fairly isotropic. However, the most likely PTA sources (pairs of supermassive black holes at the centers of merging galaxies) are discrete point sources at specific sky locations. Even if they are distributed via a discrete Poisson process, their apparent angular locations or intensities exhibit correlations. These can occur at the largest angular scales (for example dipole anisotropies [7,46] due to our motion with respect to the average Hubble flow), or they may be at much smaller angular scales.

If PTA sources are distributed in the same way as galaxies, then the relevant quantity is the power spectrum of matter density perturbations, which peaks at a distance scale of about 70 Mpc. However, it appears that even at the peak of the spectrum, these correlations are overwhelmed by "shot noise" arising from the discreteness of the individual PTA sources [47] (Fig. 3). The shot noise produces angular covariance coefficients $C_l$ which are $l$-independent for $l \lesssim l_{\max}$. Here, $l_{\max} \approx 100$ Mpc$/10$ pc $\approx 10^7$ is the ratio of the distance to the closest PTA sources to the characteristic GW wavelength.

To model and understand the effects of these correlations, we construct ensembles of cosmological models in which each realization breaks rotational invariance, but for which the full ensemble is rotationally invariant and thus has no preferred directions.

### A. Modeling angular correlations among sources: A collection of Gaussian subensembles

One way to do this is to create an ensemble of Gaussian ensembles. To avoid confusion, we will say that the full ensemble is made up of Gaussian subensembles. In this construction, each of the Gaussian subensembles breaks rotational invariance, but the full ensemble contains all rotated versions of each subensemble, and thus is rotationally invariant. While the full ensemble is no longer Gaussian [31], the key calculational methods can still be used. To compute ensemble averages, we first average over a given Gaussian subensemble, and then average over all subensembles.

Each Gaussian subensemble is constructed as in Sec. VII, but replacing the second moments given in (7.4) with

$$\langle h(f,\Omega)h^*(f',\Omega')\rangle = 2H(f)\delta(f-f')\delta^2(\Omega,\Omega')\psi(\Omega),$$
$$\langle h(f,\Omega)h(f',\Omega')\rangle = 0. \quad (9.1)$$

Here, $\psi(\Omega)$ is a real non-negative dimensionless function of the GW source direction, which describes the anisotropic distribution of GW sources within any particular Gaussian subensemble.

Averages within a given subensemble are computed as in previous sections of this paper, using Isserlis's theorem if and as needed. The angle brackets without subscripts $\langle \rangle$ in (9.1) refer to an average *only* over the subensemble labeled by $\psi$. If $\psi$ is not a constant function, then this subensemble breaks rotational invariance [48]. In contrast, averages over the full ensemble, which includes many different choices of $\psi$, will be written with a subscript as $\langle \rangle_\psi$.

We assume that the full ensemble is described by a set of functions $\psi$ whose first and second moments are given by

$$1 = \langle \psi(\Omega) \rangle_\psi, \quad (9.2)$$

$$C(\hat{\Omega} \cdot \hat{\Omega}') \equiv \langle \psi(\Omega) \psi(\Omega') \rangle_\psi - \langle \psi(\Omega) \rangle_\psi \langle \psi(\Omega') \rangle_\psi. \quad (9.3)$$

Here, the angle brackets $\langle \rangle_\psi$ denote a full ensemble average, but in practice we only use this to carry out the final average over all subensembles.

We note that knowledge of the first and second moments alone is not enough information for us to compute the ensemble average of any functional. Since only the first and second moments of $\psi$ are known, we can only compute ensemble averages of quantities that are linear or quadratic in $\psi$. Since this includes the mean and variance of the HD correlation, it is sufficient for our purposes. However, we have no equivalent of Isserlis's theorem to compute higher moments—although each subensemble is Gaussian, our full ensemble is not Gaussian [31].

Because the first moment (9.2) is independent of direction and the second moment (9.3) only depends upon the angle between $\Omega$ and $\Omega'$, the full ensemble has no preferred directions [49]. Because the first moment of $\psi$ is normalized to unity, any quantity linear in $H$ has the same expectation value as previously calculated. Thus, the normalization and interpretation of the spectral function $H(f)$ is unchanged.

The function $C(\hat{\Omega} \cdot \hat{\Omega}')$ describes the power spectrum of angular fluctuations in the GW background energy density. Using a standard normalization convention (see final paragraph of this section) it can be written as a sum of Legendre polynomials

$$C(\hat{\Omega} \cdot \hat{\Omega}') = C(\cos \beta) = \sum_{L=0}^{\infty} \frac{2L+1}{4\pi} C_L P_L(\cos \beta), \quad (9.4)$$

where, as before, $\cos \beta = \hat{\Omega} \cdot \hat{\Omega}'$. The expansion coefficients $C_L$ are constrained by (9.2), so they cannot have arbitrary values. For example, the sum of $(2L+1)C_L$ is non-negative, because

$$\langle [\psi(\Omega) - \langle \psi(\Omega) \rangle]^2 \rangle_\psi \geq 0 \Longrightarrow$$
$$\langle \psi(\Omega) \psi(\Omega) \rangle_\psi - \langle \psi(\Omega) \rangle_\psi^2 \geq 0 \Longrightarrow$$
$$C(\hat{\Omega} \cdot \hat{\Omega}) \geq 0 \Longrightarrow$$
$$C(1) \geq 0 \Longrightarrow$$
$$\sum_L (2L+1) C_L \geq 0. \quad (9.5)$$

The first inequality holds because the mean value of a non-negative quantity is non-negative, the second from completing the square, the third from (9.3), the fourth from $\hat{\Omega} \cdot \hat{\Omega} = 1$, and the fifth follows from (9.4) and $P_l(1) = 1$.

Furthermore, $\psi(\Omega) \geq 0$ implies that $C(\beta) \geq -1$ for any angle $\beta$.

Our ensemble definition is quite general, so it can be used to model different effects. For example, suppose we want to construct an ensemble of universes which is like the standard Gaussian ensemble, but for which the power spectrum $H(f)$ varies in overall amplitude from one subensemble to the next. The first moment normalization (9.2) implies that $H(f)$ is the average power spectrum of the complete ensemble (taking into account GW sources in all directions). If each subensemble has exactly that power spectrum, then $C(\hat{\Omega} \cdot \hat{\Omega}') = 0$, so the $C_l$ vanish for all $l$, and we recover the standard Gaussian ensemble. Alternatively, we can construct an ensemble with the same angular properties as the standard Gaussian ensemble, but for which the power spectrum varies in overall amplitude by a factor of 1 about $H(f)$. To obtain this, set $C(\hat{\Omega} \cdot \hat{\Omega}') = 1$, corresponding to $C_0 = 4\pi$ and $C_l = 0$ for $l > 0$. We return to this after (9.10).

Note that there is an alternative approach, which we did not follow. For that, each function $\psi$ is decomposed into spherical harmonics, $\psi(\Omega) = \sum_{lm} \psi_{lm} Y_{lm}(\Omega)$. The set of $\psi$ used to define the ensemble is then specified via a set of complex coefficients $\psi_{lm} = (-1)^m \psi^*_{l,-m}$. The properties in (9.2) are equivalent to $\langle \psi_{00} \rangle_\psi = \sqrt{4\pi}$ and $\langle \psi_{lm} \rangle_\psi = 0$ for $l > 0$, and those of (9.3) are equivalent to $\langle \psi_{lm} \psi^*_{l'm'} \rangle_\psi - \langle \psi_{lm} \rangle_\psi \langle \psi^*_{l'm'} \rangle_\psi = C_l \delta_{ll'} \delta_{mm'}$. This approach is equivalent to ours, and provides another way to understand the normalization conventions. Those employing it should beware that the $\psi_{lm}$ cannot be a set of Gaussian random variables with the above first and second moments, because those would not satisfy $\psi(\Omega) \geq 0$ for every representative function $\psi(\Omega)$ in the ensemble.

### B. Cosmic (co)variance for the ensemble with correlated source locations

Using this computational framework, we now compute the cosmic variance and covariance for the ensemble with correlated source locations. We do these quantities first, because they are considerably easier to obtain than the total variance and covariance. Those are computed later in this paper, starting in Sec. IX C.

For an extended discussion of the differences between total and cosmic (co)variance, please see [27].

Our starting point is the pulsar-averaged redshift correlation $\Gamma(\gamma)$ given by (7.7) for any realization of the universe. The average of $\Gamma(\gamma)$ over a Gaussian subensemble follows immediately from computing the expected value of (7.7) using (9.1). This simply inserts $\psi(\Omega)$ into (7.8), giving

$$\langle \Gamma(\gamma) \rangle = \frac{1}{2} h^2 \int \frac{d\Omega}{4\pi} \psi(\Omega) \Big( \mu(\gamma, \Omega, \Omega) + \mu(\gamma, \bar{\Omega}, \bar{\Omega}) \Big)$$
$$= h^2 \mu_u(\gamma) \int \frac{d\Omega}{4\pi} \psi(\Omega). \quad (9.6)$$

To obtain the second equality, we have removed the two-point function $\mu(\gamma, \Omega, \Omega)$ from the integral, since when the two points are coincident, it is independent of $\Omega$ and equal to the HD curve $\mu_u(\gamma)$.

To obtain the expected value of $\Gamma(\gamma)$ for the full ensemble, we average (9.6) over $\psi$, using the first moment (9.2). This gives

$$\langle \Gamma(\gamma) \rangle_\psi = h^2 \mu_u(\gamma), \quad (9.7)$$

which is in agreement with the isotropic result.

To find the cosmic variance and covariance, we need to compute the second moment of $\Gamma$. For a given subensemble, we carry out the same calculation which led to (7.13) in Sec. VII. We obtain

$$\langle \Gamma(\gamma)\Gamma(\gamma') \rangle = \langle \Gamma(\gamma) \rangle \langle \Gamma(\gamma') \rangle + \hbar^4 \int \frac{d\Omega}{4\pi} \int \frac{d\Omega'}{4\pi} \psi(\Omega)\psi(\Omega') \Big( \mu(\gamma, \Omega, \Omega')\mu(\gamma', \Omega', \Omega) + \mu(\bar\gamma, \Omega, \bar\Omega')\mu(\bar\gamma', \bar\Omega', \Omega) \Big)$$
$$= \int \frac{d\Omega}{4\pi} \int \frac{d\Omega'}{4\pi} \psi(\Omega)\psi(\Omega') \Big[ h^4 \mu_u(\gamma)\mu_u(\gamma') + \hbar^4 \Big( \mu(\gamma,\beta)\mu(\gamma',\beta) + \mu(\bar\gamma,\bar\beta)\mu(\bar\gamma',\bar\beta) \Big) \Big], \quad (9.8)$$

where $\cos\beta = \hat\Omega \cdot \hat\Omega'$, $\bar\beta = \pi - \beta$, $\bar\gamma = \pi - \gamma$, and $\bar\gamma' = \pi - \gamma'$. The first equality follows by repeating the calculation leading to (7.13) (the only change is that two factors of $\psi$ appear), and the second equality follows from (9.6) and the cancellation of the phase of the two-point function (6.8).

We now average (9.8) over the full ensemble using the second moment (9.3). This gives

$$\langle \Gamma(\gamma)\Gamma(\gamma') \rangle_\psi = \frac{1}{2} \int_0^\pi \sin\beta \, d\beta \Big( C(\cos\beta) + 1 \Big) \Big[ h^4 \mu_u(\gamma)\mu_u(\gamma') + \hbar^4 \Big( \mu(\gamma,\beta)\mu(\gamma',\beta) + \mu(\bar\gamma,\bar\beta)\mu(\bar\gamma',\bar\beta) \Big) \Big]. \quad (9.9)$$

To find the covariance of the full ensemble, we subtract $\langle \Gamma(\gamma) \rangle_\psi \langle \Gamma(\gamma') \rangle_\psi$, which is obtained from (9.7). Using (7.17) and (9.4), this gives the cosmic covariance for an ensemble with correlated sources, as

$$\sigma_{\cos}^2(\gamma,\gamma') = 2\hbar^4 \widetilde\mu^2(\gamma,\gamma') + \frac{C_0}{4\pi} h^4 \mu_u(\gamma)\mu_u(\gamma') + \frac{1}{2}\hbar^4 \int_0^\pi \sin\beta \, d\beta \, C(\cos\beta) \Big[ \mu(\gamma,\beta)\mu(\gamma',\beta) + \mu(\bar\gamma,\bar\beta)\mu(\bar\gamma',\bar\beta) \Big]. \quad (9.10)$$

As $C \to 0$, only the first term on the rhs survives, consistent with (7.16). Below, we simplify the $C \neq 0$ case by evaluating the integral and writing the result in terms of the coefficients $C_l$.

Before doing so, we examine the special case discussed after (9.5): setting $C(\cos\beta) = \text{constant} > 0$. This corresponds to introducing variations in the overall normalization of the GW background power spectrum $H(f)$. With this choice, different realizations of the universe have the same shape for $H(f)$, but with differing overall scales. Setting $C(\cos\beta)$ to a positive constant corresponds to $C_0 > 0$ and $C_{l>0} = 0$. Using (7.18), this shifts the covariance (9.10) by an amount $C_0[h^4 \mu_u(\gamma)\mu_u(\gamma') + 2\hbar^4 \widetilde\mu^2(\gamma,\gamma')]/4\pi$. Setting $\gamma = \gamma'$, one can see that the variance increases, compared to the standard Gaussian ensemble with $C_0 = 0$.

Returning to the general case, the integrals over $\beta$ can be evaluated by using the harmonic decomposition (6.9) of the two-point function $\mu(\gamma,\beta)$. For this, we need the integral of three Jacobi polynomials, which can be written in terms of Clebsch-Gordon coefficients or Wigner 3j symbols [50]. Since the first Jacobi polynomial is a normal Legendre polynomial, the integral that we need is

$$\int_{-1}^1 dz \left(\frac{1+z}{2}\right)^4 P_L(z) P_{l-2}^{(0,4)}(z) P_{l'-2}^{(0,4)}(z)$$
$$= 2 \begin{pmatrix} L & l & l' \\ 0 & 2 & -2 \end{pmatrix}^2. \quad (9.11)$$

Note that the rhs is symmetric in $l$ and $l'$. Letting $z = \cos\beta$ and using (6.9), this implies that

$$\int_0^\pi \sin\beta \, d\beta \, P_L(\cos\beta) \mu(\gamma,\beta) \mu(\gamma',\beta)$$
$$= 2 \sum_l \sum_{l'} a_l a_{l'} \begin{pmatrix} L & l & l' \\ 0 & 2 & -2 \end{pmatrix}^2 P_l(\cos\gamma) P_{l'}(\cos\gamma'). \quad (9.12)$$

To exploit this, we return to (9.10), replacing $C(\cos\beta)$ with its harmonic sum (9.4).

The final form of the cosmic covariance for the ensemble of correlated sky location sources follows immediately, and is

$$\sigma_{\cos}^2(\gamma,\gamma') \equiv \langle \Delta\Gamma(\gamma) \Delta\Gamma(\gamma') \rangle_\psi$$
$$= 2\hbar^4 \widetilde\mu^2(\gamma,\gamma') + \frac{C_0}{4\pi} h^4 \mu_u(\gamma) \mu_u(\gamma')$$
$$+ \hbar^4 \sum_L \frac{2L+1}{4\pi} C_L \sum_l \sum_{l'} a_l a_{l'} \begin{pmatrix} L & l & l' \\ 0 & 2 & -2 \end{pmatrix}^2$$
$$\times [1 + (-1)^{l+l'+L}] P_l(\cos\gamma) P_{l'}(\cos\gamma'). \quad (9.13)$$

The power of $-1$ arises from the $\mu(\bar\gamma,\bar\beta)\mu(\bar\gamma',\bar\beta)$ term of (9.10), because

$$P_L(\cos\bar{\beta}) = (-1)^L P_L(\cos\beta),$$
$$P_l(\cos\bar{\gamma}) = (-1)^l P_l(\cos\gamma),$$
$$P_{l'}(\cos\bar{\gamma}') = (-1)^{l'} P_{l'}(\cos\gamma'). \quad (9.14)$$

In the limit $C(\cos\beta) \to 0$, which is equivalent to $C_L \to 0$, we recover the cosmic covariance (8.4) which was computed for uncorrelated sources.

Note that the $L = 0$ term in the sum, which is proportional to $\widetilde{\mu}^2(\gamma, \gamma')$, is easily recovered from (9.13). This is because, for $m \geq 0$

$$\begin{pmatrix} 0 & l & l' \\ 0 & m & -m \end{pmatrix} = \begin{cases} 0 & \text{if } l < m \\ \frac{(-1)^{l+m}}{\sqrt{2l+1}}\delta_{ll'} & \text{if } l \geq m \end{cases}, \quad (9.15)$$

which immediately leads to (8.4).

### C. Mean and second moment of the HD correlation for a subensemble

To apply the "ensemble of Gaussian subensembles" computational framework, we need the first and second moments of the (total, not pulsar averaged) HD correlation in a given subensemble. The calculations are similar to the ones carried out in Sec. VII for the cosmic variance, but differ in one important way: the total covariance is affected by the pulsar terms, which drop out of the cosmic variance.

The pulsar term changes the frequency-independent redshift response $F(\Omega, \Omega_p)$ given in (2.2) to a frequency-dependent response, given by

$$R(fT_p, \Omega, \Omega_p) = \mathcal{T}(fT_p, \hat{\Omega}\cdot\hat{\Omega}_p) F(\Omega, \Omega_p). \quad (9.16)$$

Here, the frequency-dependent term is

$$\mathcal{T}(fT_p, \hat{\Omega}\cdot\hat{\Omega}_p) = 1 - e^{-2\pi i f T_p(1+\hat{\Omega}\cdot\hat{\Omega}_p)}, \quad (9.17)$$

where $T_p$ is the light travel time from pulsar $p$ to Earth. In (9.17), the first term is the Earth term, and the second term is the pulsar term, which can add constructively or destructively to the Earth term. Detailed derivations may be found in [5] [Eq. (32)] or [27] [Eq. (C17)].

The correlation $\rho_{pq}$ between pulsars $p$ and $q$ for any universe in any subensemble is defined by (7.5), and given by an expression similar to (7.7):

$$\rho_{pq} = \frac{1}{4}\int df \int df' \int d\Omega \int d\Omega' \,\text{sinc}(\pi(f-f')T)\Big[$$
$$h(f,\Omega)\,h^*(f',\Omega')R^*(fT_p,\Omega,\Omega_p)R(f'T_q,\Omega',\Omega_q)+$$
$$h(f,\Omega)\,h(f',\Omega')\,R^*(fT_p,\Omega,\Omega_p)R^*(f'T_q,\Omega',\Omega_q)+$$
$$h^*(f,\Omega)h^*(f',\Omega')R(fT_p,\Omega,\Omega_p)\,R(f'T_q,\Omega',\Omega_q)+$$
$$h^*(f,\Omega)h(f',\Omega')\,R(fT_p,\Omega,\Omega_p)\,R^*(f'T_q,\Omega',\Omega_q)\Big]. \quad (9.18)$$

The subensemble average of (9.18), for a given anisotropy $\psi(\Omega)$, is obtained by using (9.1). Only the first and last terms of (9.18) survive, introducing a delta function of frequency and a delta function on the sphere. Integrating out those delta functions gives the subensemble average

$$\langle\rho_{pq}\rangle = \frac{1}{2}\int df\,H(f)\int d\Omega\,\psi(\Omega)\Big[\mathcal{T}^*(fT_p,\hat{\Omega}\cdot\hat{\Omega}_p)\mathcal{T}(fT_q,\hat{\Omega}\cdot\hat{\Omega}_q)F^*(\Omega,\Omega_p)F(\Omega,\Omega_q) +$$
$$\mathcal{T}(fT_p,\hat{\Omega}\cdot\hat{\Omega}_p)\mathcal{T}^*(fT_q,\hat{\Omega}\cdot\hat{\Omega}_q)F(\Omega,\Omega_p)F^*(\Omega,\Omega_q)\Big]. \quad (9.19)$$

By making reasonable assumptions about the pulsar distances and sky locations, this expression can be further simplified.

Consider the integration over $\Omega$ in (9.19). The products $\mathcal{T}^*\mathcal{T}$ consist of a slowly varying part and a rapidly varying part. We make two assumptions. (A) That $\psi(\Omega)$ only varies on angular scales greater than $\approx 1/fT_{\text{pulsar}} \approx T/T_{\text{pulsar}} \approx 10^{-2}$ radians, where $T_{\text{pulsar}}$ is a typical Earth-pulsar or pulsar-pulsar light propagation time, and $T$ is the total observation time. (B) That $T_{\text{pulsar}}$ is larger than the characteristic coherence time of the GW background (see Eq. (C13) in [27]). Then, as discussed before Eq. (45) of [5], the rapidly varying terms integrate to zero. The slowly varying terms in the products $\mathcal{T}^*\mathcal{T}$ are 1 if $p$ and $q$ denote distinct pulsars, and are 2 if $p$ and $q$ denote the same pulsar, so

$$\mathcal{T}^*(fT_p,\hat{\Omega}\cdot\hat{\Omega}_p)\mathcal{T}(fT_q,\hat{\Omega}\cdot\hat{\Omega}_q) \to 1 + \delta_{pq}. \quad (9.20)$$

This doubling arises through the autocorrelation of the pulsar term, and is discussed in more detail below.

Assumption (A) about the angular scale of variation of $\psi$, which corresponds to $l$ values of a few hundred, is sufficient but not necessary. The pulsar terms will also integrate to zero for $p \neq q$ if the variations in $\psi$ are uncorrelated with the pulsar locations. The average over the full ensemble (which we construct to be rotationally invariant) should guarantee that this condition is satisfied for nonpathological choices of $\psi$.

With these assumptions, the correlation between pulsars in any Gaussian subensemble is given by

$$\langle \rho_{pq} \rangle = h^2 (1 + \delta_{pq}) \int \frac{d\Omega}{4\pi} \psi(\Omega) \varrho_{pq}(\Omega), \qquad (9.21)$$

where the squared strain $h^2$ is defined by (7.9), and $\varrho_{pq}(\Omega)$ is defined by (2.13). The effect of the $1 + \delta_{pq}$ factor is to double the correlation of a pulsar with itself, in comparison with the correlation between two pulsars that lie along the same line of sight separated by a distance larger than a GW wavelength; a detailed discussion may be found between Eqs. (C11) and (C13) of [27] and around Eq. (45) of [5].

For later use, we now compute the second moment of $\rho_{pq}$ for a particular Gaussian subensemble, starting from (9.18). The calculation is very similar to the one which leads to (7.13). The result is

$$\langle \rho_{pq} \rho_{rs} \rangle = \langle \rho_{pq} \rangle \langle \rho_{rs} \rangle + \frac{\hbar^4}{h^4} \left[ \langle \rho_{pr} \rangle \langle \rho_{qs} \rangle + \langle \rho_{ps} \rangle \langle \rho_{qr} \rangle \right], \qquad (9.22)$$

where $\hbar^4$ is given in (7.14).

It is not surprising that (9.22) takes exactly the same form as [38] [Eq. (2.11)], since the average is over a single Gaussian subensemble. Since (9.21) is linear in $\psi$, the rhs of (9.22) is quadratic in $\psi$. Hence, the average of (9.22) over different subensembles can be computed by employing (9.3).

### D. The mean and total covariance (of the HD correlation) for the full ensemble

To obtain the mean and total covariance of the HD correlation, we average over the different Gaussian subensembles. This average over different $\psi(\Omega)$ yields the full ensemble average.

The mean correlation between pulsars $p$ and $q$ is obtained by averaging (9.21) over the different subensembles using (9.2). This gives the full ensemble average

$$\begin{aligned}\langle \rho_{pq} \rangle_\psi &\equiv \langle\!\langle \rho_{pq} \rangle\!\rangle_\psi \\ &= h^2 (1 + \delta_{pq}) \int \frac{d\Omega}{4\pi} \varrho_{pq}(\Omega) \\ &= h^2 (1 + \delta_{pq}) \mu_{\mathrm{u}}(\gamma_{pq}) \\ &= h^2 \mu_{pq}, \end{aligned} \qquad (9.23)$$

where $\gamma_{pq}$ is the sky separation angle between pulsars $p$ and $q$. The first equality in (9.23) indicates the average over all subensembles, the second equality follows from (9.2), which sets $\psi \to 1$ in (9.21), and from (3.3), which shows that the $F^*F$ and $FF^*$ terms both average to the same real quantity. The third equality follows from (3.3), which defines the unpolarized HD curve $\mu_{\mathrm{u}}(\gamma)$. In the final line, we have adopted the notation of [38] [Eqs. (2.3) and (2.4)] and defined the HD correlation matrix

$$\mu_{pq} \equiv (1 + \delta_{pq}) \mu_{\mathrm{u}}(\gamma_{pq}). \qquad (9.24)$$

The entries in this matrix of HD correlations are obtained from the values of the HD curve at the angular separations of pulsars $p$ and $q$, apart from the entries along the diagonal. Those have been doubled by the pulsar term contribution to the autocorrelation. Thus, (9.23) implies that the mean of HD correlation for the full ensemble is identical to that found in the standard Gaussian ensemble: both are proportional to the HD curve.

The total covariance of $\rho_{pq}$ is easily obtained for the full ensemble. To compute this, it is helpful to first compute the ensemble average over $\psi$ of $\langle \rho_{pq} \rangle \langle \rho_{rs} \rangle$. We replace $\langle \rho_{pq} \rangle$ and $\langle \rho_{rs} \rangle$ with (9.21), and then use (9.3) and (9.23) to compute the ensemble average over $\psi$. Since $\langle \psi(\Omega) \psi(\Omega') \rangle_\psi = C(\hat{\Omega} \cdot \hat{\Omega}') + 1$, this gives

$$\langle\!\langle \rho_{pq} \rangle \langle \rho_{rs} \rangle\!\rangle_\psi = h^4 (\mu_{pq} \mu_{rs} + \mathcal{D}_{pq,rs}), \qquad (9.25)$$

where $\mu_{pq}$ is the HD correlation matrix given in (9.24), and we have defined

$$\mathcal{D}_{pq,rs} \equiv (1 + \delta_{pq})(1 + \delta_{rs}) D_{pq,rs}, \qquad (9.26)$$

where

$$D_{pq,rs} \equiv \int \frac{d\Omega}{4\pi} \int \frac{d\Omega'}{4\pi} C(\hat{\Omega} \cdot \hat{\Omega}') \varrho_{pq}(\Omega) \varrho_{rs}(\Omega'). \qquad (9.27)$$

In all of these equations, $p$, $q$, $r$ and $s$ label pulsars, any or all of which could be distinct or identical.

The full ensemble average $\langle \rho_{pq} \rho_{rs} \rangle_\psi \equiv \langle\!\langle \rho_{pq} \rho_{rs} \rangle\!\rangle_\psi$ is obtained from (9.22) by using (9.25) to average the three terms over $\psi$. This gives

$$\begin{aligned}\langle \rho_{pq} \rho_{rs} \rangle_\psi &= h^4 (\mu_{pq} \mu_{rs} + \mathcal{D}_{pq,rs}) + \hbar^4 (\mu_{pr} \mu_{qs} \\ &\quad + \mu_{ps} \mu_{qr} + \mathcal{D}_{pr,qs} + \mathcal{D}_{ps,qr}).\end{aligned} \qquad (9.28)$$

The covariance of the full ensemble is obtained by evaluating $\langle \rho_{pq} \rangle_\psi \langle \rho_{rs} \rangle_\psi$ with (9.23), and then subtracting it from (9.28). This eliminates the first term on the rhs of (9.28), giving the total covariance

$$\begin{aligned}\mathcal{C}_{pq,rs} &\equiv \langle \rho_{pq} \rho_{rs} \rangle_\psi - \langle \rho_{pq} \rangle_\psi \langle \rho_{rs} \rangle_\psi \\ &= \hbar^4 (\mu_{pr} \mu_{qs} + \mu_{ps} \mu_{qr}) \\ &\quad + h^4 \mathcal{D}_{pq,rs} + \hbar^4 (\mathcal{D}_{pr,qs} + \mathcal{D}_{ps,qr}).\end{aligned} \qquad (9.29)$$

This is one of our paper's main results, since we will now derive explicit formulas for the different terms.

If the GW source locations in the ensemble are uncorrelated and all Gaussian subensembles have the same overall GW intensity, then $\mathcal{D}_{pq,rs} \to 0$. The covariance in (9.29) then reduces to the terms on the first line of the final equality. This is the covariance of the standard Gaussian ensemble, as given in [38] [Eq. (2.10)]. It only depends upon $\hbar$ and not upon $h$.

If the GW source locations in the ensemble are uncorrelated but the GW intensity varies between subensembles (meaning $C_l = 0$ for $l > 0$ and $C_0 \neq 0$) then the covariance also depends upon $h$. In this case $\mathcal{D}_{pq,rs} \to (C_0/4\pi)\mu_{pq}\mu_{rs}$, so that

$$\mathcal{C}_{pq,rs} = \left(1 + \frac{C_0}{4\pi}\right)\hbar^4(\mu_{pr}\mu_{qs} + \mu_{ps}\mu_{qr})$$
$$+ \frac{C_0}{4\pi}h^4\mu_{pq}\mu_{rs}. \quad (9.30)$$

If the GW background is broadband (in the sense of [27] [Eq. (C31)], where the ratio $\hbar^4/h^4 \propto 1/T \to 0$ as the observation time $T \to \infty$) then the $h^4$ term dominates, reflecting the overall variation in GW intensity among different subensembles.

The covariance is useful in several contexts. For example, to reconstruct the HD correlation from experimental data, the relative weighting of correlations within a given angular bin (in $\gamma$) is determined from the covariance [38] [Eq. (3.10)]. Here, we evaluate the covariance and variance in closed form.

To compute the covariance, we first express the HD integrand (2.13) as a sum of spherical harmonics

$$\varrho_{pq}(\Omega) = \sum_{lm} P_{lm}(\Omega_p, \Omega_q) Y^*_{lm}(\Omega). \quad (9.31)$$

Here, the amplitudes are

$$P_{lm}(\Omega_p, \Omega_q) \equiv \frac{A_{lm}(\Omega_p, \Omega_q) + A_{lm}(\Omega_q, \Omega_p)}{2}, \quad (9.32)$$

where

$$A_{lm}(\Omega_p, \Omega_q) \equiv \int d\Omega F(\Omega, \Omega_p) F^*(\Omega, \Omega_q) Y_{lm}(\Omega). \quad (9.33)$$

The two terms in (9.32) arise from taking the real part of $F(\Omega, \Omega_p)F^*(\Omega, \Omega_q)$ as shown in the second equality of (2.13).

From the Legendre polynomial expansion (9.4) of $C(\hat{\Omega} \cdot \hat{\Omega}')$ and the addition theorem (3.4), it follows from (9.27) and (9.31) that

$$D_{pq,rs} = \frac{1}{16\pi^2}\sum_{LM} C_L P_{LM}(\Omega_p, \Omega_q) P^*_{LM}(\Omega_r, \Omega_s). \quad (9.34)$$

The harmonic amplitudes $P_{lm}(\Omega_p, \Omega_q)$ were first studied in [7]. There, they were computed in "position space" for small $l \leq 2$. Then, they were obtained for all $l$ in Appendix E of [8]. In both cases, special pulsar positions were used, with $p$ at the North Pole of the sphere and $q$ along the line of longitude $\phi = 0$. Here, we provide an (infinite harmonic sum) expression which is valid for *any* pulsar pair.

We evaluate $A_{lm}(\Omega_p, \Omega_q)$ by substituting the diagonal form (2.10) for $F$ into (9.33) twice, and integrating over $\Omega$. The integrand is a product of three spherical harmonics, with spin weights 0, 2 and $-2$ [the $-2$ arises from complex conjugation, see (A2)]. Using (A7), this may be written in terms of Wigner 3j symbols, as

$$A_{lm}(\Omega_p, \Omega_q) = \sum_{l_1 m_1}\sum_{l_2 m_2} A_{l_1} A_{l_2} Y^*_{l_1 m_1}(\Omega_p) Y_{l_2 m_2}(\Omega_q)$$
$$\times \int d\Omega\, Y_{lm}(\Omega)\, _2Y_{l_1 m_1}(\Omega)\, _2Y^*_{l_2 m_2}(\Omega)$$
$$= \sum_{l_1 m_1}\sum_{l_2 m_2} A_{l_1} A_{l_2} Y^*_{l_1 m_1}(\Omega_p) Y_{l_2 m_2}(\Omega_q)$$
$$\times \sqrt{\frac{(2l+1)(2l_1+1)(2l_2+1)}{4\pi}}$$
$$\times (-1)^{m_2}\begin{pmatrix} l & l_1 & l_2 \\ 0 & -2 & 2 \end{pmatrix}\begin{pmatrix} l & l_1 & l_2 \\ m & m_1 & -m_2 \end{pmatrix}. \quad (9.35)$$

The final equality of (A4) implies that the summand vanishes unless $m_2 = m + m_1$, so the double sum over $m_1$ and $m_2$ may be rewritten as a single sum.

Combining the $A_{lm}$ according to (9.32) gives the spherical harmonic coefficients

$$P_{lm}(\Omega_p, \Omega_q) = \sum_{l_1 m_1}\sum_{l_2 m_2} A_{l_1} A_{l_2} Y^*_{l_1 m_1}(\Omega_p) Y^*_{l_2 m_2}(\Omega_q)$$
$$\times \sqrt{\frac{(2l+1)(2l_1+1)(2l_2+1)}{16\pi}}$$
$$\times \left[1 + (-1)^{l+l_1+l_2}\right]\begin{pmatrix} l & l_1 & l_2 \\ 0 & -2 & 2 \end{pmatrix}$$
$$\times \begin{pmatrix} l & l_1 & l_2 \\ m & m_1 & m_2 \end{pmatrix}, \quad (9.36)$$

where we have used (A2) to write $Y_{l_2 m_2}(\Omega_q) = (-1)^{m_2}Y^*_{l_2,-m_2}(\Omega_q)$ and flipped the sign of $m_2$, to obtain an expression that is explicitly symmetric under interchange of pulsars $p$ and $q$. This is because changing the sign of the second row of either of the Wigner 3j symbols introduces a factor of $(-1)^{l+l_1+l_2}$, see (A4).

Note that if $l = m = 0$, then by virtue of (9.15), only the diagonal terms $l_1 = l_2$ and $m_1 = m_2$ survive in expressions (9.35) and (9.36). These reduce to the sum in (3.3), giving $P_{00} = A_{00} = \sqrt{4\pi}\mu_u(\gamma)$.

Combining these results provides an explicit expression for $D_{pq,rs}$, from which the covariance matrix for any sky positions may be obtained. Substituting (9.36) into (9.34) gives

$$D_{pq,rs} = \sum_L \frac{2L+1}{256\pi^3} C_L \sum_{l_1,\ldots,l_4} A_{l_1} A_{l_2} A_{l_3} A_{l_4}$$
$$\times \sqrt{(2l_1+1)(2l_2+1)(2l_3+1)(2l_4+1)}$$
$$\times \left[1+(-1)^{L+l_1+l_2}\right]\left[1+(-1)^{L+l_3+l_4}\right]$$
$$\times \begin{pmatrix} L & l_1 & l_2 \\ 0 & -2 & 2 \end{pmatrix} \begin{pmatrix} L & l_3 & l_4 \\ 0 & -2 & 2 \end{pmatrix}$$
$$\times G_{Ll_1l_2l_3l_4}(\Omega_p, \Omega_q, \Omega_r, \Omega_s), \quad (9.37)$$

where the rotationally invariant function of the pulsar sky positions is

$$G_{Ll_1l_2l_3l_4}(\Omega_p, \Omega_q, \Omega_r, \Omega_s)$$
$$= \sum_{M=-L}^{L} \sum_{m_1=-l_1}^{l_1} \sum_{m_2=-l_2}^{l_2} \sum_{m_3=-l_3}^{l_3} \sum_{m_4=-l_4}^{l_4}$$
$$\times \begin{pmatrix} L & l_1 & l_2 \\ M & m_1 & m_2 \end{pmatrix} \begin{pmatrix} L & l_3 & l_4 \\ M & m_3 & m_4 \end{pmatrix}$$
$$\times Y^*_{l_1 m_1}(\Omega_p) Y^*_{l_2 m_2}(\Omega_q) Y_{l_3 m_3}(\Omega_r) Y_{l_4 m_4}(\Omega_s). \quad (9.38)$$

It should be possible to express $G_{Ll_1l_2l_3l_4}$ as a function of Legendre polynomials of the dot products $\hat{\Omega}_p \cdot \hat{\Omega}_r$, $\Omega_p \cdot \hat{\Omega}_s$, $\hat{\Omega}_q \cdot \hat{\Omega}_r$, and $\hat{\Omega}_q \cdot \hat{\Omega}_s$.

### E. The total Hellings and Downs variance for the full ensemble

The total variance $\sigma_{\text{tot}}^2$ is obtained from the covariance $\mathcal{C}_{pq,rs}$ in (9.29) by setting $r \to p$ and $s \to q$. From rotational invariance, $\sigma_{\text{tot}}^2$ is only a function of the angle $\gamma$ between pulsars $p$ and $q$, with $\cos\gamma = \hat{\Omega}_p \cdot \hat{\Omega}_q$. So, from (9.26) and (9.29), we obtain

$$\sigma_{\text{tot}}^2(\gamma) = \mathcal{C}_{pq,pq}$$
$$= \hbar^4(\mu_{pp}\mu_{qq} + u_{pq}^2) + h^4(1+\delta_{pq})^2 D_{pq,pq}$$
$$+ \hbar^4\big((1+\delta_{pp})(1+\delta_{qq})D_{pp,qq} + (1+\delta_{pq})^2 D_{pq,pq}\big)$$
$$= \hbar^4\big[(1+3\delta_{pq})\mu_u^2(\gamma) + 4\mu_u^2(0) + 4D_{pp,qq}(\gamma)\big]$$
$$+ (h^4 + \hbar^4)(1+3\delta_{pq})D_{pq,pq}(\gamma). \quad (9.39)$$

For the third equality, we have used (9.24) and $\delta_{pq}^2 = \delta_{pq}$, and explicitly indicated the dependence of $D$ on the angle $\gamma$. To determine this completely, we return to (9.37) and evaluate $D_{pq,pq}(\gamma)$ and $D_{pp,qq}(\gamma)$ as sums of Legendre polynomials in $\gamma$. The results may be found in (9.43), (9.47), and (9.52).

Since $D_{pq,pq}$ is only a function of $\gamma$, it can be pulsar averaged without changing its value. The required quantity is the pulsar average $\langle G_{Ll_1l_2l_3l_4}(\Omega_p, \Omega_q, \Omega_p, \Omega_q)\rangle_{pq \in \gamma}$. The pulsar average of the four spherical harmonics which appear in (9.38) is

$$\langle Y^*_{l_1 m_1}(\Omega_p) Y^*_{l_2 m_2}(\Omega_q) Y_{l_3 m_3}(\Omega_p) Y_{l_4 m_4}(\Omega_q)\rangle_{pq \in \gamma}$$
$$= \frac{1}{4\pi} \sum_{lm} P_l(\cos\gamma) \int d\Omega_p Y_{lm}(\Omega_p) Y^*_{l_1 m_1}(\Omega_p) Y_{l_3 m_3}(\Omega_p) \int d\Omega_q Y^*_{lm}(\Omega_q) Y^*_{l_2 m_2}(\Omega_q) Y_{l_4 m_4}(\Omega_q)$$
$$= \frac{1}{16\pi^2} \sum_{lm} (2l+1) P_l(\cos\gamma) \sqrt{(2l_1+1)(2l_2+1)(2l_3+1)(2l_4+1)} (-1)^{m+m_1+m_2}$$
$$\times \begin{pmatrix} l & l_1 & l_3 \\ 0 & 0 & 0 \end{pmatrix} \begin{pmatrix} l & l_2 & l_4 \\ 0 & 0 & 0 \end{pmatrix} \begin{pmatrix} l & l_1 & l_3 \\ m & -m_1 & m_3 \end{pmatrix} \begin{pmatrix} l & l_2 & l_4 \\ -m & -m_2 & m_4 \end{pmatrix}. \quad (9.40)$$

The first equality follows from the recipe (4.4) for pulsar averaging, and the second equality from the standard formula (A7) for the integral of three spin-weighted spherical harmonics.

This immediately gives the total variance $D_{pq,pq}$. Inserting (9.40) into (9.38) and then inserting (9.38) into (9.37) yields

$$D_{pq,pq}(\gamma) = \langle D_{pq,pq}\rangle_{pq \in \gamma}$$
$$= \sum_{LM} \sum_{lm} \sum_{l_1 m_1} \sum_{l_2 m_2} \sum_{l_3 m_3} \sum_{l_4 m_4} \frac{(2L+1)(2l+1)}{8\pi} C_L s_{l_1} s_{l_2} s_{l_3} s_{l_4} P_l(\cos\gamma)(-1)^{M+m}\left[1+(-1)^{L+l_3+l_4}\right]$$
$$\times \begin{pmatrix} L & l_1 & l_2 \\ 0 & -2 & 2 \end{pmatrix} \begin{pmatrix} L & l_3 & l_4 \\ 0 & -2 & 2 \end{pmatrix} \begin{pmatrix} l & l_1 & l_3 \\ 0 & 0 & 0 \end{pmatrix} \begin{pmatrix} l & l_2 & l_4 \\ 0 & 0 & 0 \end{pmatrix} \begin{pmatrix} L & l_1 & l_2 \\ M & m_1 & m_2 \end{pmatrix} \begin{pmatrix} L & l_3 & l_4 \\ M & m_3 & m_4 \end{pmatrix}$$
$$\times \begin{pmatrix} l & l_1 & l_3 \\ m & -m_1 & m_3 \end{pmatrix} \begin{pmatrix} l & l_2 & l_4 \\ -m & -m_2 & m_4 \end{pmatrix}. \quad (9.41)$$

For this, we have defined constants

$$s_l \equiv \frac{2l+1}{4\pi}(-1)^l A_l$$
$$= \begin{cases} 0 & \text{for } l < 2 \\ \frac{2l+1}{\sqrt{(l+2)(l+1)l(l-1)}} & \text{for } l \geq 2 \end{cases} \quad (9.42)$$

to simplify the appearance of (9.41) and subsequent equations. Note that the sign disappears, since the summand of (9.41) vanishes unless $l_1 + l_2 + l_3 + l_4$ is even.

To obtain (9.41), we made several simplifications. First, we replaced $(-1)^{m_1+m_2}$ with $(-1)^M$, because nonvanishing terms must have the bottom row of each Wigner 3j symbol sum to zero (A4), implying that $M = -m_1 - m_2$. Second, we used

$$\frac{1}{2}\left[1+(-1)^{L+l_1+l_2}\right]\left[1+(-1)^{L+l_3+l_4}\right] = 1+(-1)^{L+l_3+l_3},$$

since the only nonzero terms in the sum have $l_1 + l_2 + l_3 + l_4$ even. This is because a Wigner 3j symbol vanishes if the bottom row vanishes and the sum of the top row is odd. Hence $l + l_1 + l_3$ and $l + l_2 + l_4$ are both even, implying that $l_1 + l_2 + l_3 + l_4$ is even.

Expression (9.41) provides a convenient decomposition of the total variance into a sum of Legendre polynomials of $\cos\gamma$, where $\gamma$ is the angle between the directions to pulsars $p$ and $q$. For this purpose, define a matrix of coefficients $d_{Ll}$ via

$$D_{pq,pq}(\gamma) = \sum_L \sum_l d_{Ll} C_L P_l(\cos\gamma). \quad (9.43)$$

There are two alternative approaches which allow further simplifications in the formula for the coefficients $d_{Ll}$.

In the first approach, note that the Wigner 3j symbols in (9.41) vanish if the sum of the bottom row is nonzero (A4). This means that the summation over $M, m, m_1, m_2, m_3, m_4$ can be replaced by a summation over $M, m, m_1$, with $m_2 = -M - m_1$, $m_3 = m_1 - m$ and $m_4 = m - M - m_1$. This gives

$$d_{Ll} = \sum_{M=-L}^{L}\sum_{m=-l}^{l}\sum_{l_1 m_1}\sum_{l_2}\sum_{l_3}\sum_{l_4}(-1)^{M+m}\frac{(2l+1)(2L+1)}{8\pi}\left[1+(-1)^{L+l_3+l_4}\right]$$
$$\times s_{l_1}s_{l_2}s_{l_3}s_{l_4}\begin{pmatrix}L & l_1 & l_2 \\ 0 & -2 & 2\end{pmatrix}\begin{pmatrix}L & l_3 & l_4 \\ 0 & -2 & 2\end{pmatrix}\begin{pmatrix}l & l_1 & l_3 \\ 0 & 0 & 0\end{pmatrix}\begin{pmatrix}l & l_2 & l_4 \\ 0 & 0 & 0\end{pmatrix}\begin{pmatrix}L & l_1 & l_2 \\ M & m_1 & -M-m_1\end{pmatrix}$$
$$\times \begin{pmatrix}L & l_3 & l_4 \\ M & m_1-m & m-M-m_1\end{pmatrix}\begin{pmatrix}l & l_1 & l_3 \\ m & -m_1 & m_1-m\end{pmatrix}\begin{pmatrix}l & l_2 & l_4 \\ -m & M+m_1 & m-M-m_1\end{pmatrix}, \quad (9.44)$$

which has four infinite sums over $l_1, l_2, l_3, l_4$ and three finite sums over $M, m,$ and $m_1$.

A simpler and more symmetric expression can be obtained by returning to (9.41) and using the Wigner $6j$ symbol to carry out the sums over $M, m, m_1, m_2, m_3, m_4$. The Wigner $6j$ symbol satisfies the equation

$$\begin{Bmatrix}j_1 & j_2 & j_3 \\ j_4 & j_5 & j_6\end{Bmatrix} = \sum_{n_1,\ldots,n_6}(-1)^{\sum_{k=1}^{6}(j_k-n_k)}\begin{pmatrix}j_1 & j_2 & j_3 \\ -n_1 & -n_2 & -n_3\end{pmatrix}\begin{pmatrix}j_1 & j_5 & j_6 \\ n_1 & -n_5 & n_6\end{pmatrix}\begin{pmatrix}j_4 & j_2 & j_6 \\ n_4 & n_2 & -n_6\end{pmatrix}\begin{pmatrix}j_4 & j_5 & j_3 \\ -n_4 & n_5 & n_3\end{pmatrix}. \quad (9.45)$$

We make the following substitutions into (9.45):

$$j_1 = l_1, \quad n_1 = -m_1, \quad j_2 = l_2, \quad n_2 = -m_2,$$
$$j_3 = L, \quad n_3 = -M, \quad j_4 = l_4, \quad n_4 = m_4,$$
$$j_5 = l_3, \quad n_5 = -m_3, \quad j_6 = l, \quad n_6 = m. \quad (9.46)$$

Then, we exploit properties of the Wigner 3j symbol (A4). Swapping any pair of columns or inverting the signs of the bottom row multiplies the Wigner 3j symbol by $(-1)^S$, where $S$ denote the sum of the top row. Using these, we arrive at

$$d_{Ll} = \frac{(-1)^{L+l}}{8\pi}(2l+1)(2L+1)\sum_{l_1,l_2,l_3,l_4}\left[1+(-1)^{L+l_3+l_4}\right]s_{l_1}s_{l_2}s_{l_3}s_{l_4}\begin{pmatrix}l & l_1 & l_3 \\ 0 & 0 & 0\end{pmatrix}\begin{pmatrix}l & l_2 & l_4 \\ 0 & 0 & 0\end{pmatrix}\begin{pmatrix}L & l_1 & l_2 \\ 0 & -2 & 2\end{pmatrix}$$
$$\times \begin{pmatrix}L & l_3 & l_4 \\ 0 & -2 & 2\end{pmatrix}\begin{Bmatrix}l_1 & l_2 & L \\ l_4 & l_3 & l\end{Bmatrix}. \quad (9.47)$$

This formula for the numerical coefficients is very pretty, and it may be possible to cancel some terms in this sum by exploiting further symmetries of the Wigner 3j and 6j symbols.

To complete the evaluation of the total variance (9.39), we also need to evaluate evaluate $D_{pp,qq}(\gamma)$. Return to the definition (9.27), where from (2.13) and (2.7) the HD integrand is

$$\varrho_{pp}(\Omega) = F(\Omega, \Omega_p) F^*(\Omega, \Omega_p) = \frac{1}{4}(1 - \hat{\Omega} \cdot \hat{\Omega}_p)^2, \quad (9.48)$$

and the final equality follows from (2.7). If we let $z = \hat{\Omega} \cdot \hat{\Omega}_p$, then

$$\begin{aligned}
\varrho_{pp}(\Omega) &= \frac{1}{4}z^2 - \frac{1}{2}z + \frac{1}{4} \\
&= \frac{1}{6}P_2(z) - \frac{1}{2}P_1(z) + \frac{1}{3}P_0(z) \\
&= \sum_{m=-2}^{2} \frac{2\pi}{15} Y_{2m}(\Omega_p) Y^*_{2m}(\Omega) \\
&- \sum_{m=-1}^{1} \frac{2\pi}{3} Y_{1m}(\Omega_p) Y^*_{1m}(\Omega) \\
&+ \frac{4\pi}{3} Y_{00}(\Omega_p) Y^*_{00}(\Omega).
\end{aligned} \quad (9.49)$$

For the second equality, we have expressed the quadratic polynomial in terms of Legendre polynomials, and for the third equality, we have used the addition theorem (3.4) for $l = 0$, 1, and 2. Thus, the expansion coefficients $P_{lm}(\Omega_p, \Omega_p)$ given in (9.32) are

$$P_{lm}(\Omega_p, \Omega_p) = \begin{cases} \frac{4\pi}{3} Y_{lm}(\Omega_p) & \text{if } l = 0 \text{ and } m = 0, \\ -\frac{2\pi}{3} Y_{lm}(\Omega_p) & \text{if } l = 1 \text{ and } |m| \leq 1, \\ \frac{2\pi}{15} Y_{lm}(\Omega_p) & \text{if } l = 2 \text{ and } |m| \leq 2, \\ 0 & \text{otherwise.} \end{cases} \quad (9.50)$$

Corresponding expressions for $P_{lm}(\Omega_q, \Omega_q)$ are obtained by setting $\Omega_p \to \Omega_q$ in (9.50).

We now complete the evaluation of $D_{pp,qq}(\gamma)$. From (9.34), the quantity required is

$$\sum_M P_{LM}(\Omega_p, \Omega_p) P^*_{LM}(\Omega_q, \Omega_q) = \begin{cases} \frac{4\pi}{9} P_0(\hat{\Omega}_p \cdot \hat{\Omega}_q) & \text{if } L=0, \\ \frac{\pi}{3} P_1(\hat{\Omega}_p \cdot \hat{\Omega}_q) & \text{if } L=1, \\ \frac{\pi}{45} P_2(\hat{\Omega}_p \cdot \hat{\Omega}_q) & \text{if } L=2, \\ 0 & \text{if } L>2, \end{cases} \quad (9.51)$$

where the values are taken from (9.50) with corresponding expressions with $\Omega_p \to \Omega_q$, and the sum over $M$ is done via the addition theorem (3.4). Substituting (9.51) into (9.34) gives

$$D_{pp,qq}(\gamma) = \frac{C_0}{36\pi} P_0(\cos\gamma) + \frac{C_1}{48\pi} P_1(\cos\gamma) + \frac{C_2}{720\pi} P_2(\cos\gamma), \quad (9.52)$$

with $\cos\gamma = \hat{\Omega}_p \cdot \hat{\Omega}_q$. Substituting this and $D_{pq,pq}(\gamma)$ as defined by (9.43) and (9.47) into (9.39) gives the total variance of the HD correlation.

## X. CONCLUSION

We have shown how harmonic analysis, based on the diagonal decomposition (2.10), makes it straightforward to calculate the most important quantities of interest for pulsar timing arrays. We then use these methods to model universes whose GW source sky positions have nontrivial angular correlations. To do this modeling, we build "statistically isotropic ensembles" from anisotropic Gaussian subensembles. This leads to simple equations for the cosmic variance/covariance, and for the total variance/covariance. Investigations for realistic cosmological models are underway [51], though for large $l$ these effects may be too small to be observable in the near future.

*Note added.* Recently, we learned that Agarwal and Romano had independently carried out the calculation of the cosmic variance for the ensemble with nontrivial angular correlations [52]. Their results are consistent with those obtained in Sec. IX B of this paper; in fact we have unified our notation so that the results may be easily compared. Subsequently, we applied these methods to estimate the impact of galaxy discreteness and clustering on the HD correlations [47].

## ACKNOWLEDGMENTS

I would like to thank Joe Romano and Neil Cornish for helpful discussions, Nastassia Grimm for persuading me that it makes sense to construct an ensemble of Gaussian subensembles, Serena Valtolina for checking some of the calculations and discovering that (9.44) could be written in terms of the Wigner 6j symbol (9.47), Daniel Pook-Kolb for numerically checking (2.10) and computing a table of $d_{Ll}$ values, Marc Favata for confirming that (A7) can be violated if $s_1 + s_2 + s_3 \neq 0$, Neha Anil Kumar for pointing out an equation equivalent to (2.10) in Sec. V.B of [14], and Joe Romano and Deepali Agarwal for comparing results and converging notation, and Gabi Richardson for assistance in manuscript preparation. I also thank the anonymous referee for useful suggestions and for identifying several oversights and omissions.

## APPENDIX A: SPIN-WEIGHTED SPERICAL HARMONICS

For convenience, we list a few of the key formulas for spin-weighted spherical harmonics. These are reproduced from the complete listing given in [8] (Appendix A).

### 1. Spin weight zero

$$Y_{lm}(\Omega) \equiv {}_0Y_{lm}(\Omega). \tag{A1}$$

Throughout this paper, we drop the prefix "0" from the spin-0 weighted harmonics, which are the conventional spherical harmonics.

### 2. Complex conjugation

$${}_sY^*_{lm}(\Omega) = (-1)^{m+s}{}_{-s}Y_{l,-m}(\Omega). \tag{A2}$$

### 3. Inversion on the sphere (also called parity)

$${}_sY_{lm}(\bar{\Omega}) = (-1)^l{}_{-s}Y_{lm}(\Omega)$$
$$= (-1)^{m+s+l}{}_sY^*_{l,-m}(\Omega)$$
$$\Omega = (\theta, \phi) \Leftrightarrow \bar{\Omega} = (\pi - \theta, \phi + \pi). \tag{A3}$$

### 4. Symmetries/properties of the Wigner 3j symbol

$$\begin{pmatrix} l_1 & l_2 & l_3 \\ m_1 & m_2 & m_3 \end{pmatrix} = (-1)^{l_1+l_2+l_3}\begin{pmatrix} l_2 & l_1 & l_3 \\ m_2 & m_1 & m_3 \end{pmatrix}$$
$$= (-1)^{l_1+l_2+l_3}\begin{pmatrix} l_1 & l_3 & l_2 \\ m_1 & m_3 & m_2 \end{pmatrix}$$
$$= (-1)^{l_1+l_2+l_3}\begin{pmatrix} l_1 & l_2 & l_2 \\ -m_1 & -m_2 & -m_3 \end{pmatrix}$$
$$= 0 \quad \text{if } m_1 + m_2 + m_3 \neq 0. \tag{A4}$$

Hence, the symbol is invariant under (a) any even permutation of columns or (b) any odd permutation of the columns accompanied by a sign flip of the bottom row.

### 5. Spin-2 harmonics used in this paper

These vanish for $l < 2$ and may be obtained for $l \geq 2$ by taking derivatives of the normal (spin-weight 0) spherical harmonics:

$${}_2Y_{lm}(\theta, \phi) \equiv \sqrt{\frac{(l-2)!}{(l+2)!}} \eth_1\eth_0 Y_{lm}(\theta, \phi), \tag{A5}$$

where the "edth" spin-raising operators are

$$\eth_s \equiv -(\sin\theta)^s\left(\frac{\partial}{\partial\theta} + \frac{i}{\sin\theta}\frac{\partial}{\partial\phi}\right)(\sin\theta)^{-s}. \tag{A6}$$

### 6. Integral of three spherical harmonics

If $s_1 + s_2 + s_3 = 0$, then

$$\int d\Omega \, {}_{s_1}Y_{l_1m_1}(\Omega) \, {}_{s_2}Y_{l_2m_2}(\Omega) \, {}_{s_3}Y_{l_3m_3}(\Omega) = \sqrt{\frac{(2l_1+1)(2l_2+1)(2l_3+1)}{4\pi}}\begin{pmatrix} l_1 & l_2 & l_3 \\ m_1 & m_2 & m_3 \end{pmatrix}\begin{pmatrix} l_1 & l_2 & l_3 \\ -s_1 & -s_2 & -s_3 \end{pmatrix}. \tag{A7}$$

Note: the condition $s_1 + s_2 + s_3 = 0$ was omitted from Eq. (A13) of [8]. If $s_1 + s_2 + s_3 \neq 0$, then (A7) may not hold: the lhs may be nonzero, but the rhs vanishes. For example, $\int d\Omega \, {}_0Y_{00}(\Omega)\, {}_0Y_{22}(\Omega)\, {}_2Y_{2,-2}(\Omega) = \sqrt{3/32\pi}$, whereas the rhs of (A7) vanishes for $s_1 = s_2 = 0$ and $s_3 = 2$. In such cases, the integral may be evaluated using the method of [53] (Appendix A).

## APPENDIX B: DERIVATION OF THE DIAGONAL FORM OF $F(\Omega, \Omega_p)$

Here, we derive the diagonal form of $F(\Omega, \Omega_p)$ given in (2.10), following an approach inspired by [8] (Sec. III. D). We also explain how (2.10) can be checked/verified directly, by explicitly carrying out the sums. Lastly, we perform two simple sanity checks.

To verify the diagonal form in (2.10) directly, use the addition theorem for spin-weighted harmonics [8] [Eqs. (A9)–(A11) with $s = 2, s' = 0$] to carry out the sum over $m$. Then, use

$${}_2Y_{l0}(\theta, \phi) = \sqrt{\frac{2l+1}{4\pi}}\sqrt{\frac{(l-2)!}{(l+2)!}}P_l^2(\cos\theta) \tag{B1}$$

and

$$\frac{1}{2}(1-z) = \sum_{l=2}^{\infty}\frac{(-1)^l(2l+1)}{(l+2)(l+1)l(l-1)}P_l^2(z) \tag{B2}$$

[derived in [9] Eq. (42)] to complete the sum over $l$. Some algebra with trigonometric identities leads directly to (2.2).

[In (B1) and (B2), $P_l^m(z)$ denotes an associated Legendre function, *not* the square of a Legendre polynomial.]

To derive the diagonal form in (2.10), begin with $F(\hat{z}, \Omega_p)$ as given in (2.8). This is the response of a pulsar at an arbitrary sky direction $\hat{\Omega}_p$ to a GW with direction $\hat{z}$. We rotate this pattern to obtain the response to a GW with arbitrary direction $\hat{\Omega}$. (Here, and in what follows, it is often helpful to write the arguments of $F$ and spherical harmonics as unit vectors rather than as coordinates on the sphere.)

There are many different rotations that will bring $\hat{z}$ to $\hat{\Omega}$. For the reasons explained in Sec. II, we select the unique rotation that *consistently maintains the directions of the polarization vectors $\hat{m}$ and $\hat{n}$, as defined by* (2.3).

Rotations are defined by three Euler angles [54] [Eqs. (3.35)–(3.37)] conventionally denoted $\alpha$, $\beta$ and $\gamma$, corresponding to rotation matrices

$$R(\alpha, \beta, \gamma) \equiv \begin{bmatrix} \cos\gamma & \sin\gamma & 0 \\ -\sin\gamma & \cos\gamma & 0 \\ 0 & 0 & 1 \end{bmatrix} \begin{bmatrix} \cos\beta & 0 & -\sin\beta \\ 0 & 1 & 0 \\ \sin\beta & 0 & \cos\beta \end{bmatrix} \begin{bmatrix} \cos\alpha & \sin\alpha & 0 \\ -\sin\alpha & \cos\alpha & 0 \\ 0 & 0 & 1 \end{bmatrix}. \quad (B3)$$

This matrix acts from the left, on column vectors whose three entries are the $\hat{x}$, $\hat{y}$ and $\hat{z}$ components. (In this appendix, $\gamma$ and $\beta$ denote rotation angles. Elsewhere in the paper, they denote the angles between pairs of pulsars or between pairs of GW sources.)

By inspection, the rotation in (B3) acting on $\hat{z}$ gives

$$R(\alpha, \beta, \gamma)\hat{z} = -\cos\gamma\sin\beta\hat{x} + \sin\gamma\sin\beta\hat{y} + \cos\beta\hat{z}. \quad (B4)$$

Thus, to obtain the GW direction $\hat{\Omega} = R(\alpha, \beta, \gamma)\hat{z}$ as given in (2.1), we must set $\beta = -\theta$ and $\gamma = -\phi$. Note that $\alpha$ can take any value. This is also obvious from inspection of (B3), since the rightmost matrix leaves $\hat{z}$ invariant.

However, there is only a *single* value of $\alpha$ which yields the correct polarization vectors $\hat{m}$ and $\hat{n}$, as given in (2.3). To see this, act on $\hat{n}(\hat{z})$ with the rotation $R(\alpha, -\theta, -\phi)$. The $\hat{z}$ component of $R(\alpha, -\theta, -\phi)\hat{n}$ is $\sin\theta\sin(\phi - \alpha)$. Since $\hat{n}(\hat{\Omega})$ has no $\hat{z}$ component, we must have $\alpha = \phi + N\pi$ for $N$ integer. Only even $N$, equivalent to $N = 0$, maintains the orientation of $\hat{n}$. Thus, the only acceptable rotation which carries $\hat{z}$ to $\hat{\Omega}$ *and* which carries $\hat{n}(\hat{z})$ to $\hat{n}(\hat{\Omega})$ is

$$R = R(\phi, -\theta, -\phi). \quad (B5)$$

This rotation matrix also carries $\hat{m}(\hat{z})$ to $\hat{m}(\hat{\Omega})$.

We emphasize this point one last time. For an arbitrary rotation $\mathcal{R}$, $F(\mathcal{R}\hat{\Omega}, \mathcal{R}\hat{\Omega}_p) \neq F(\hat{\Omega}, \hat{\Omega}_p)$. Equality is only obtained for rotations that satisfy $\mathcal{R}\hat{m}(\hat{\Omega}) = \hat{m}(\mathcal{R}\hat{\Omega})$ and $\mathcal{R}\hat{n}(\hat{\Omega}) = \hat{n}(\mathcal{R}\hat{\Omega})$. In words: the pulsar response $F$ is only invariant under simultaneous rotations of the GW source and pulsar directions which *also* preserve the polarization vectors $\hat{m}$ and $\hat{n}$. This consistency requirement has also been noted by others [14] [Footnote 4].

From here, it is straightforward. We first express the unrotated response function (2.8) as a sum of spherical harmonics

$$F(\hat{z}, \Omega_p) = \frac{1}{2}(1 - \cos\theta_p)e^{2i(\phi_p - \phi)}$$
$$= \sum_l q_l Y^*_{l,-2}(\Omega_p), \quad (B6)$$

with expansion coefficients $q_l$. Because the $\phi_p$ dependence in the first equality is $e^{2i\phi_p}$, the sum only includes spherical harmonics with $m = -2$, which implies that the $q_l$ vanish if $l < 2$. For $l \geq 2$ they are

$$q_l = (-1)^l \sqrt{\frac{4\pi(2l+1)}{(l+2)(l+1)l(l-1)}} e^{-2i\phi}, \quad (B7)$$

which follows immediately from (B2).

Next, we rotate the response function, by rotating the spherical harmonics. Since the rotation matrix (B5) preserves the polarization directions, rotational invariance implies that

$$F(R\hat{z}, R\hat{\Omega}_p) = F(\hat{z}, \hat{\Omega}_p) = \sum_l q_l Y^*_{l,-2}(\Omega_p), \quad (B8)$$

where the final equality comes from (B6). Setting $R\hat{z} = \hat{\Omega}$ in (B8), and then noting that, since the equation holds for all $\hat{\Omega}_p$, we can send $\hat{\Omega}_p \to R^{-1}\hat{\Omega}_p$, we obtain

$$F(\Omega, \Omega_p) = \sum_l q_l Y^*_{l,-2}(R^{-1}\hat{\Omega}_p)$$
$$= \sum_l q_l \sum_{m=-l}^{l} [D^l_{m,-2}(R^{-1}) Y_{lm}(\Omega_p)]^*$$
$$= \sum_{lm} q_l [D^l_{m,-2}(R^{-1})]^* Y^*_{lm}(\Omega_p), \quad (B9)$$

where $D^l_{mm'}$ is the Wigner D-matrix. (For fixed $l$, the $Y_{lm}$ form a $2l + 1$-dimensional vector space representation of the group $SO(3)$. Thus, the rotated $Y_{l,-2}$ is a sum of harmonics with the same $l$ and all allowed $m$ values [42] [Pg. 51].

The second equality of Eq. (B9) is obtained using

$$Y_{lm}(R\Omega) = \sum_{m'} D^l_{m'm}(R) Y_{lm'}(\Omega) \quad (B10)$$

[54] [Eq. (16.52)], which is consistent with our choice of Euler angles in (B3) and with [46] [Eqs. (7.3)–(7.7)]. Note that the corresponding relationship in [42] [Eqs. (2.43) and (2.45)] replaces $D^l_{m,-2}(R^{-1})$ in (B9) with $D^l_{m,-2}(R)$. This is equivalent: since [42] uses active rather than passive rotations, the signs of the Euler angles and their ordering are inverted, swapping $R$ and $R^{-1}$, see [42] [Eq. (1.54)] and [55] [Eq. (6.39)].

The inverse of the rotation matrix (B5) can be found by inspection of (B3), and is $R^{-1} = R(\phi, \theta, -\phi)$. This rotation carries $\hat{\Omega}$ to $\hat{z}$, while also preserving the polarization vectors.

The complex conjugate of the Wigner D-matrix is

$$[D^l_{m,-2}(R^{-1})]^* = \sqrt{\frac{4\pi}{2+1}} \,_2 Y_{lm}(\theta, \phi) e^{2i\phi}. \quad (B11)$$

This is obtained from the second line of [8] [Eq. (A6)] by setting $\phi \to \phi$, $\theta \to \theta$, $\psi \to -\phi$, $m \to -2$, and $m' \to m$. Substituting (B7) and (B11) into (B9) immediately gives the desired diagonal form (2.10).

The reader might find it helpful to carry out two simple sanity checks. First, verify (2.10) for $\hat{\Omega} = \hat{z}$. One can easily see that (2.8) follows from

$$_2 Y_{lm}(\theta = 0, \phi) = \sqrt{\frac{2l+1}{4\pi}} e^{-2i\phi} \delta_{m,-2} \quad (B12)$$

and (B2). A second simple check is to set $\hat{\Omega}_p = \hat{z}$ in (2.10). Then,

$$Y_{lm}(\theta = 0, \phi) = \sqrt{\frac{2l+1}{4\pi}} \delta_{m,0} \quad (B13)$$

and (B1) should be used. Together with (B2), they imply that $F(\Omega, \hat{z}) = (1 - \cos\theta)/2$.

## APPENDIX C: LINEAR POLARIZATION COMPONENTS OF THE TWO-POINT FUNCTION

Some calculations (see [27] for examples) are best carried out using two-point functions for linear polarization components, written $\mu_{++}$, $\mu_{\times\times}$, $\mu_{\times+}$, and $\mu_{+\times}$. Here, we extract these from the complex two-point function $\mu(\gamma, \Omega, \Omega_p)$.

These two-point functions are real, and are defined by the pulsar averages

$$\mu_{++}(\gamma, \Omega, \Omega') \equiv \langle F_+(\Omega, \Omega_p) F_+(\Omega', \Omega_q) \rangle_{pq \in \gamma} \quad (C1)$$

and corresponding pulsar averages for the other combinations of linear polarizations. The polarization components are the real and imaginary parts of the response: $F_+(\Omega, \Omega_p) \equiv \Re F(\Omega, \Omega_p)$ and $F_\times(\Omega, \Omega_p) \equiv \Im F(\Omega, \Omega_p)$, as discussed in the text following (2.4) and (2.5). The real two-point functions such as (C1) should be compared to the complex $\mu(\gamma, \Omega, \Omega')$ defined by (6.1) and explicitly calculated in (6.8) and (6.9).

We start by computing $\mu_{++}(\gamma, \Omega, \Omega')$. Using (C1) and taking the real part of $F$, it is

$$\begin{aligned}
\mu_{++}(\gamma, \Omega, \Omega') &\equiv \langle F_+(\Omega, \Omega_p) F_+(\Omega', \Omega_q) \rangle_{pq \in \gamma} \\
&= \frac{1}{4} \langle [F(\Omega, \Omega_p) + F^*(\Omega, \Omega_p)] \\
&\quad \times [F(\Omega', \Omega_q) + F^*(\Omega', \Omega_q)] \rangle_{pq \in \gamma} \\
&= \frac{1}{4} [\mu(\gamma, \Omega, \Omega') + \mu(\gamma, \bar{\Omega}, \bar{\Omega}') \\
&\quad + \mu(\bar{\gamma}, \bar{\Omega}, \Omega') + \mu(\bar{\gamma}, \Omega, \bar{\Omega}')] \\
&= \frac{1}{2} \mu(\gamma, \beta) \cos 2\chi + \frac{1}{2} \mu(\bar{\gamma}, \bar{\beta}) \cos 2\bar{\chi}. \quad (C2)
\end{aligned}$$

The second equality follows from the definition of $F_+$, the third from (2.12) and (6.1), and the final equality from (6.8).

To simplify notation, we have defined

$$\chi \equiv \chi(\Omega, \Omega') = -\chi(\Omega', \Omega)$$
$$= \chi(\bar{\Omega}', \bar{\Omega}) = -\chi(\bar{\Omega}, \bar{\Omega}') \quad (C3)$$

and

$$\begin{aligned}
\bar{\chi} &\equiv \bar{\chi}(\Omega, \Omega') \equiv \chi(\Omega, \bar{\Omega}') \\
&= -\chi(\bar{\Omega}, \Omega') = \chi(\Omega', \bar{\Omega}) = \bar{\chi}(\Omega', \Omega). \quad (C4)
\end{aligned}$$

Note that $\chi$ is an antisymmetric function of its arguments, whereas $\bar{\chi}$ is a symmetric function of its arguments.

The reflection properties of Legendre and Jacobi polynomials provide an elegant form for $\mu(\bar{\gamma}, \bar{\beta})$, where $\bar{\gamma} = \pi - \gamma$ and $\bar{\beta} = \pi - \beta$. For the Legendre polynomials, $\cos\bar{\gamma} = \cos(\pi - \gamma) = -\cos\gamma$ and $P_l(-z) = (-1)^l P_l(z)$. For the Jacobi polynomials, $\cos\bar{\beta} = \cos(\pi - \beta) = -\cos\beta$, and $P_l^{(a,b)}(-z) = (-1)^l P_l^{(b,a)}(z)$. The transformation of the overall factor follows from $\cos(\pi/2 - \beta/2) = \sin(\beta/2)$. Using these together with (6.9) immediately gives

$$\mu(\bar{\gamma}, \bar{\beta}) = \left(\sin\frac{\beta}{2}\right)^4 \sum_l a_l P_{l-2}^{(4,0)}(\cos\beta) P_l(\cos\gamma), \quad (C5)$$

where $a_l$ are defined in (2.11), and the reader should note the reversed ordering in the upper indices of the Jacobi polynomial.

Calculations similar to (C2) for the other combinations of linear polarizations give

$$\mu_{++}(\gamma,\Omega,\Omega') = \frac{1}{2}\left[\mu(\gamma,\beta)\cos 2\chi + \mu(\bar{\gamma},\bar{\beta})\cos 2\bar{\chi}\right],$$

$$\mu_{\times\times}(\gamma,\Omega,\Omega') = \frac{1}{2}\left[\mu(\gamma,\beta)\cos 2\chi - \mu(\bar{\gamma},\bar{\beta})\cos 2\bar{\chi}\right],$$

$$\mu_{\times+}(\gamma,\Omega,\Omega') = \frac{1}{2}\left[\mu(\gamma,\beta)\sin 2\chi + \mu(\bar{\gamma},\bar{\beta})\sin 2\bar{\chi}\right],$$

$$\mu_{+\times}(\gamma,\Omega,\Omega') = \frac{1}{2}\left[-\mu(\gamma,\beta)\sin 2\chi + \mu(\bar{\gamma},\bar{\beta})\sin 2\bar{\chi}\right]. \quad \text{(C6)}$$

Since $\chi$ is an antisymmetric function of its arguments and $\bar{\chi}$ is a symmetric function of its arguments, one can see that swapping the polarization indices and the source directions leaves $\mu$ invariant: $\mu_{AA'}(\gamma,\Omega,\Omega') = \mu_{A'A}(\gamma,\Omega',\Omega)$. Hence, under interchange of the source directions, $\mu_{\times\times}$ and $\mu_{++}$ are invariant, and $\mu_{+\times}$ and $\mu_{\times+}$ are swapped.

The linear polarization components of the two-point function (C6) satisfy a relationship which is useful for some calculations [47]. If all four terms on the lhs of (C6) are squared and summed, one obtains the "gauge-independent" relationship

$$\sum_{AA'}\mu^2_{AA'}(\gamma,\Omega,\Omega') = \frac{1}{2}\left[\mu^2(\gamma,\beta) + \mu^2(\bar{\gamma},\bar{\beta})\right]. \quad \text{(C7)}$$

While the summands on the lhs depend upon the phases $\chi$ and $\bar{\chi}$, on the rhs these have canceled out: the sum only depends upon $\gamma$ and $\beta$.

The relationships in (C6) generalize Eqs. (G9) and (G10) of [27], which are computed for points $\Omega$ and $\Omega'$ that lie on the same "line of longitude." For such points, $\phi' - \phi = 0$ and $\bar{\phi}' - \phi = \pi$, so (6.5) implies that $\cos\chi = \cos\bar{\chi} = 1$ and $\sin\chi = \sin\bar{\chi} = 0$. For such points, (C6) then reduces to Eqs. (G9) and (G10) from [27].

ensuring that if the ensemble contains a Gaussian subensemble described by $\Psi(\Omega)$, then it also contains Gaussian subensembles for all rotated versions of the function $\Psi(R\Omega)$, where $R$ is a rotation matrix.